    \newcommand{\ba}{\begin{eqnarray}}
    \newcommand{\ea}{\end{eqnarray}}
    \newcommand{\be}{\begin{equation}}
    \newcommand{\ee}{\end{equation}}

    \newcommand{\IN}{\mathrm{in}}
    
    \newcommand{\OUT}{\mathrm{out}}

    \newcommand{\zlk}{\mathrm{ZLK}}

    \def\e1{e_1^2}

    \newcommand*{\EDIT}[1]{#1}

    \documentclass[twocolumn, twocolappendix]{aastex631}
    \usepackage{amsmath, graphicx}

    \submitjournal{ApJ}

\begin{document}

\shorttitle{Superthermal Eccentricity Distributions}
\title{Revisiting the Tertiary-induced Binary Black Hole Mergers: the Role of Superthermal Wide Tertiary Eccentricity Distributions}

\correspondingauthor{Yubo Su}
\email{yubosu@princeton.edu}

\author[0000-0001-8283-3425]{Yubo Su}
\affiliation{
Department of Astrophysical Sciences, Princeton University, 4 Ivy Lane, Princeton, NJ 08544, USA
}

\author[0000-0002-0643-8295]{Bin Liu}
\affiliation{Institute for Astronomy, School of Physics, Zhejiang University, 310058 Hangzhou, China\\}
\affiliation{Niels Bohr International Academy, Niels Bohr Institute, Blegdamsvej 17, 2100 Copenhagen, Denmark\\
}

\author[0000-0002-0458-7828]{Siyao Xu}
\affiliation{
Department of Physics, University of Florida, 2001 Museum Rd., Gainesville, FL 32611, USA
}

\begin{abstract}

Recent studies show that the eccentricity distribution of wide binaries
(semimajor axis $\gtrsim10^3\;\mathrm{AU}$) observed by \emph{Gaia} tends to favor large
eccentricities more strongly than the canonical thermal distribution ($P(e)
\propto e$)---such distributions are termed ``superthermal''. Motivated by this
observation, we revisit the formation channel of black hole (BH) binary mergers
in triple stellar systems and study the impact of superthermal eccentricity
distributions in the outer binaries. We explore the persistence of the highly
eccentric outer orbits after each component in a stellar triple has undergone
mass loss due to supernova explosions. We find that the outer eccentricity
distribution can remain significantly superthermal for modestly hierarchical
BH triples satisfying $a_\IN/a_\OUT\gtrsim 0.005$ (where $a_\IN$ and $a_\OUT$
are the semimajor axes of the inner and outer orbits), and are otherwise shaped
by mass-loss induced kicks and dynamical instability. We then study the impact
of these different outer eccentricity distributions of the remaining BH triples
on mergers via the tertiary-induced channel. Of interest, we find that mergers
can sometimes be produced even when the initial stellar orbits are near
alignment (not subject to the von-Zeipel-Lidov-Kozai effect; ZLK effect) as long
as the system is sufficiently hierarchical. On the other hand, although the
impact of the octupole-order ZLK effect is much greater when the outer binary is
more eccentric, we find that the merger fraction only changes modestly for
extreme outer eccentricity distributions, as the largest eccentricities tend to
lead to dynamical instability.

\end{abstract}


\section{Introduction}


The LIGO-Virgo-KAGRA Collaboration has detected about 90 double compact object merger events during the first three observing runs \citep{LIGO_O3b}.
Merging stellar-mass black hole binaries (BHB) are suggested to have diverse formation channels and environments.
These include isolated binary evolution \citep[e.g.,][]{lipunov1997black,
lipunov2017first, podsiadlowski2003formation, belczynski2010effect,
belczynski2016first, dominik2012double, dominik2013double, dominik2015double},
chemically homogeneous evolution \citep[e.g.][]{mandel2016merging, de2016chemically, marchant2016new, riley2021chemically},
and multiple-body evolution in the gas disks of active galactic nuclei
\citep[e.g.][]{mckernan2012, stone2017, leigh2018rate, secunda2019, tagawa2020formation, li2021, li2022, li2022b, samsing2022agn}.
In addition to these, two main dynamical channels are proposed, involving scenarios of intense gravitational scatterings in the dense clusters \citep[e.g.,][]{zwart1999black,
o2006binary, miller2009mergers, banerjee2010stellar, downing2010compact,
ziosi2014dynamics, rodriguez2015binary, samsing2017assembly, samsing2018black,
rodriguez2018post, gondan2018eccentric}, and tertiary-induced mergers through von Zeipel-Lidov-Kozai (ZLK) oscillations \citep[e.g.,][]{miller2002four,
wen2003eccentricity, Antonini_2012, Antonini_2014, antonini2017binary, silsbee2017lidov,
LL17, LL18, randall2018induced, randall2018analytical, hoang2018black,
fragione2019, fragione2019loeb, bin_misc5, LL19, LLW_apjl, bin_misc2, su2020spin,
liu2021hierarchical,su2021mass}.

In the scenario of ZLK-induced mergers,
we consider an inner BHB with masses $m_1$, $m_2$, semimajor axis $a_\IN$ and eccentricity $e_\IN$, and a tertiary companion ($m_3$) that moves around
the center of mass of the inner bodies on a wider orbit with $a_\OUT$ and $e_\OUT$.
When the tertiary
companion is on a sufficiently inclined (outer) orbit ($40^\circ\lesssim I_\OUT\lesssim 140^\circ$ where $I_\OUT$ is the inclination angle between the inner and outer orbital planes),
the leading order, quadrupolar gravitational interactions between the orbits results in eccentricity oscillations in the inner binary. This is known as the ZLK effect \citep[][]{zeipel, lidov, kozai}.
For sufficiently misaligned orbits ($I_\OUT \simeq 90^\circ$), the inner binary can attain extreme eccentricities, which result in enhanced gravitational wave (GW) radiation and can greatly reduce the merger time of BHBs, even inducing mergers of BHBs that would not merge in isolation.
Based on the findings of \citet{liu2021hierarchical}, all crucial elements of the ZLK-induced merger channel,
including the merger window (the range of $I_\OUT$ that permits BHB merger) and merger fraction, can be understood in an analytical way.

When the inner binary has unequal masses and the outer orbit is eccentric,
octupole-order corrections to the standard ZLK effect begin to appear \citep[e.g.,][]{ford2000secular, blaes2002kozai, lithwick2011eccentric, LML15, naoz2016eccentric}.
The strength of these effects is quantified by the dimensionless parameter
\begin{align}
    \epsilon_{\rm oct} = \frac{m_1 - m_2}{m_1 + m_2} \frac{a_\IN}{a_\OUT}
        \frac{e_{\rm out}}{1 - e_{\rm out}^2}.\label{eq:eps_oct}
\end{align}
In general, the octupole-order effects
increase the inclination window for extreme eccentricity excitation, thereby
further enhancing the rate of successful binary mergers \citep{LL18}.
Since the octupole strength, $\epsilon_{\rm oct}$, increases with decreasing binary mass ratio ($m_2/m_1$)
(see Eq.~\ref{eq:eps_oct}),
ZLK-induced mergers tend to favor binaries with smaller mass ratios.
As demonstrated in~\citet{su2021mass}, the merger fractions for hierarchical triples with semimajor axis ratio $a_\IN/a_\OUT\gtrsim0.01$
can sometimes increase by up to a factor of 20 as the mass ratio decreases from unity to 0.2.

The progenitors of BH triples are likely stellar multiples. Recent studies have
unveiled intriguing characteristics of binary/triple stars observed by
\emph{Gaia} mission. In particular, wide binaries, defined as having separations
greater than $\sim10^3\;\mathrm{AU}$, exhibit a ``superthermal'' eccentricity distribution
(for a power-law eccentricity distibution $P(e) \propto e^\alpha$, they find
$\alpha > 1$; \citealp{Toko20,hwang2022}). This could be explained by assuming
binary stars are formed in turbulent and magnetized molecular clouds: Since the
velocity dispersion of young stars is found to be scale-dependent and follows a
turbulent velocity scaling similar to that of star-forming gas
\citep{Ha21,Krol21,Zhou21,Ha22}, binary stars formed in turbulent clouds are
suggested to have extreme eccentricity distributions with significant excess at
high eccentricities \citep[e.g.][]{xu2023wide}. At the same time, \emph{Gaia}
observations reveal that a large fraction of close binaries are in triple
systems with wide tertiaries (at $10^3$--$10^4\;\mathrm{AU}$), and the distributions of the
wide tertiaries' separations and eccentricities are similar to those of wide
binaries \citep{Hwa23}. While in recent literature many mechanisms have been
proposed to form/evolve BH triples ranging from stellar evolution in the inner
binary \citep{hamers2022, toonen2022, kummer2023} to dynamical formation \citep{trani2022,
atallah2024} to galactic tides \citep{grishin2022}, we focus on the case of
isolated triples where the outer tertiary  separation is
in the superthermal eccentricity regime, i.e.\ $10^3$--$10^4\;\mathrm{AU}$.

Motivated by these recent studies, in this paper,
we revisit the scenario of ZLK-induced merger by adopting an observationally supported and physically justified superthermal eccentricity distribution for the outer orbits.
We start with hierarchical, massive stellar triples and
study in detail how these (highly) superthermal eccentricity distributions of the outer orbits can be `preserved' when the black hole (BH) triples are formed.
In particular, we are interested in how the
merger window (and fraction) induced by the ZLK channel might be modified by such distributions.
The main goal of
this paper aims to quantify the dependence of the merger fraction/probability on the
outer eccentricity distribution.
Without loss of generality, we focus on
the cases where the tertiary mass is comparable to the binary BH masses. When
the tertiary mass $m_3$ is much larger than $m_{12}=m_1+m_2$ (as in the case of
a supermassive BH tertiary), dynamical stability of the triple requires $a_{\rm
out} (1-e_{\rm out}) /[a_\IN(1+e_\IN)]  \gtrsim 3.7 (m_3/m_{12})^{1/3} \gg 1$
\citep[where $e_\IN$ and $e_\OUT$ are the inner and outer eccentricities,][]{kiseleva}, which implies that the octupole effect is negligible.

This paper is organized as follows. In Section~\ref{sec 2}, we study the effect
of mass-loss induced kick during the formation of BH triples and examine the
persistence of highly superthermal outer eccentricity distributions.
In Section~\ref{s:popsynth}, we study tertiary-induced BH mergers and calculate
the merger fraction for the survived triples that have undergone supernova (SN)
explosions. The effect of the initial extreme eccentricity distribution are
explored. We discuss our results and summarize our main findings in
Section~\ref{sec 4}.

\section{Orbital Properties of Black Hole Triples at Formation}\label{sec 2}

In this section, we focus on how the orbital parameters of a stellar triple can be modified
when a BH triple is formed. Since we only focus on wide orbits ($a_\IN$,
$a_\OUT\gg10\;\mathrm{AU}$), no
stellar evolution (i.e., stellar winds, radius expansion and mass transfer) are included in
our calculations.

\subsection{Pre-kick eccentricities \texorpdfstring{$e_\OUT^0$}{eout0}}\label{sec 2 1}

Since the BH triples we will later consider have $a_\OUT \geq
3000\;\mathrm{AU}$, the stellar progenitor triples of these systems also
generally have $a_\OUT \gtrsim 3000\;\mathrm{AU}$.
As such, we expect that the outer eccentricities $e_\OUT^0$ of the stellar progenitor systems follow a superthermal eccentricity distribution \citep{hwang2022, Hwa23}. For our study, we will use the theoretical eccentricity distribution obtained by \citet{xu2023wide}, whose results we summarize below.

\citet{xu2023wide} showed that if the initial relative velocities of wide
stellar pairs (with separations $\gtrsim 10^3\;\mathrm{AU}$) follow a turbulent velocity distribution, then
the eccentricity distribution these binaries at an initial separation $r$ is given by
\begin{equation}\label{eq:pe}
   p(e) = C e \int^{\sqrt{(1+e)/2}}_{\sqrt{(1-e)/2}}\frac{f(u)du}{u\sqrt{1-u^2}\sqrt{e^2-(2u^2-1)^2}},
\end{equation}
where
\begin{equation}
   f(u) = \sqrt{\frac{2}{\pi}} \frac{1}{\sigma_u^3}\exp\Bigg(-\frac{u^2}{2\sigma_u^2}\Bigg)u^2
\end{equation}
is the turbulent speed distribution, $e$ is the eccentricity,
$C$ is a normalization constant, $u=v/v_\text{bon}(r)$,
$v$ is the relative velocity between the binary stars,
$v_\text{bon}(r) = \sqrt{2GM/r}$
is their gravitationally bound velocity,
$M$ is their total mass,
$\sigma_u = \sigma_v /v_\text{bon}(r)$,
$\sigma_v \approx 0.7 v_\text{tur} (r)$,
\begin{equation}
   v_\text{tur} (r) = \langle v \rangle = V_L \Big(\frac{r}{L}\Big)^\frac{1}{3}
\end{equation}
is the turbulent velocity at $r$,
$V_L \sim 10~$km s$^{-1}$ is the typical turbulent velocity
at the typical injection scale
$L\sim 100~$pc for interstellar turbulence
\citep{Cham20},
and the power-law index $1/3$ comes from the
Kolmogorov scaling of turbulence, as supported by observations
\citep{Yuen22}.
For wide stellar binaries with $a \gtrsim 10^3\;\mathrm{AU}$,
$p(e)$ gives a ``highly superthermal" eccentricity distribution that is
significantly concentrated at high eccentricities (see the top panel of Fig.~\ref{fig:indicies} for an example).
The deficiency at low eccentricities is caused by the small $f(u)$ near
$u=\sqrt{1/2}$ (corresponding to a circular orbit and $e = 0$) with
$v_\text{tur}(r)$ much smaller than
$\sqrt{1/2}v_\text{bon}(r)$.

\subsection{SN explosion and mass-loss induced kick}\label{sec 2 2}

\begin{figure*}
\centering
\begin{tabular}{cccc}
\includegraphics[width=16cm]{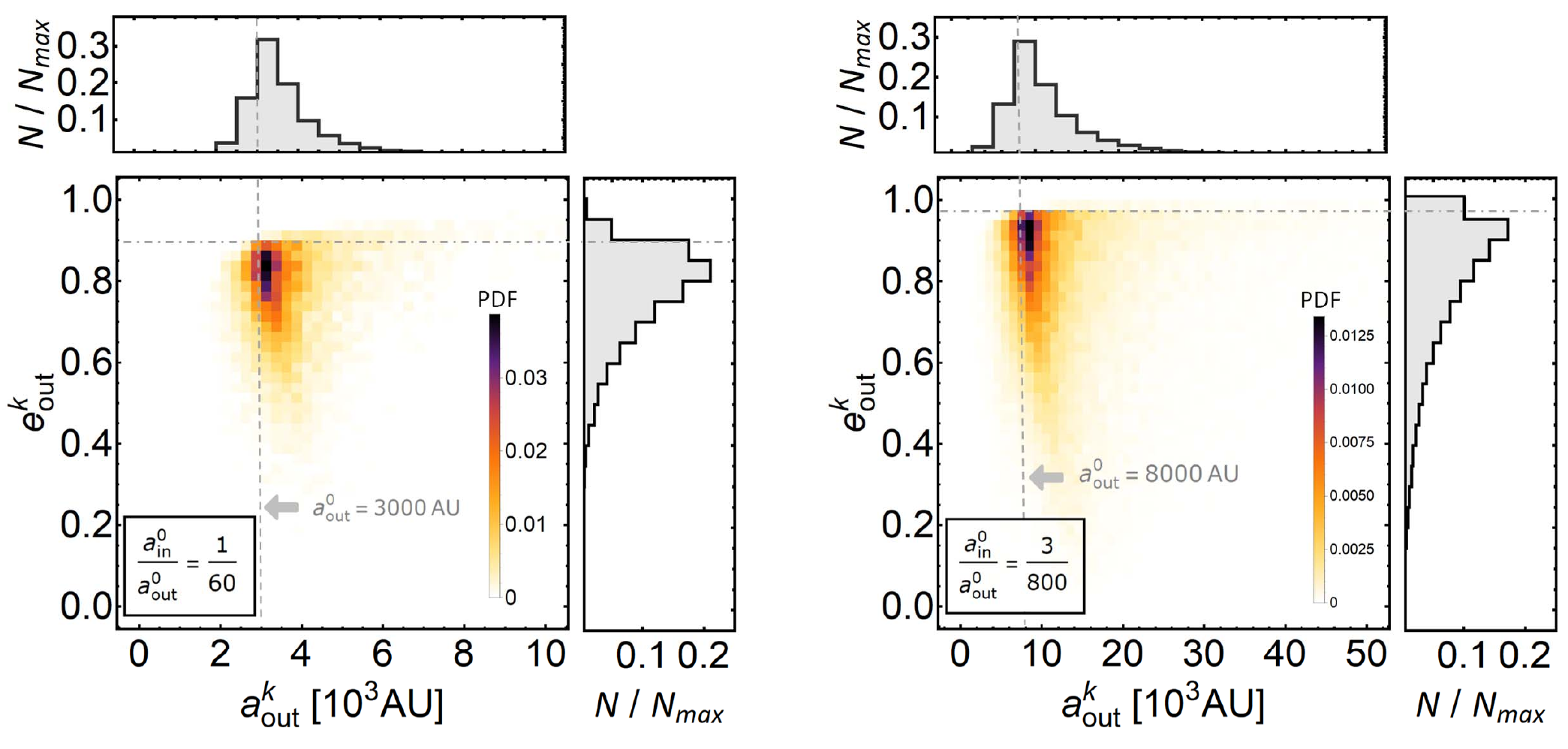}
\end{tabular}
\caption{The distributions of $a_\OUT$ and $e_\OUT$ for the survived triples
that have undergone three mass-loss induced kicks, where the left panel shows
the post-SN systems initialized with $a_\IN^0=50\;\mathrm{AU}$ and
$a_\OUT^0=3000\;\mathrm{AU}$
while the results of the right panel are from the systems with
$a_\IN^0=30\;\mathrm{AU}$
and $a_\OUT^0=8000\;\mathrm{AU}$. All of the results presented here are from stellar
triples with highly superthermal $e_\OUT^0$ distributions (Eq.~\ref{eq:pe}). The
PDF (color coded) denotes the probability density distributions of $a_\OUT^k$
and $e_\OUT^k$, which are depicted more accurately on the top and side. The
dot-dashed line in each panel corresponds to the maximum $e_\OUT^0$ that is
allowed by the stability with $I_\OUT^0=90^\circ$.}\label{fig:Post kick triples}
\end{figure*}

In our scenario, each star in a triple undergoes stellar evolution and
eventually collapses into a BH, accompanied by a sudden mass loss during the SN
explosion due to neutrino emission. The sudden change of mass induces an
impulsive change of the outer binary semimajor axis and eccentricity: we refer
to these kicks as \emph{mass-loss induced kicks}, which are also called
mass-loss kicks and Blaauw kicks \citep[e.g.,][]{blaauw1961, repetto2012}.

Note that the newly-born BH may also receive a natal kick if the neutrino
emission is anisotropic. Since the magnitude of these natal kicks on
first-generation BHs is highly uncertain, we assume the natal kick to be
negligible and only consider the mass-loss induced kick in our analysis (see
Section~\ref{ss:discussion} for discussion of the accuracy of this assumption).
If the mass-loss induced kick velocity is not too high, the inner binary may
remain bound to the tertiary star, allowing the three components to survive all
the SNs and form a stable BH triple. Subsequently, the inner binary may
eventually merge via ZLK oscillations if the mutual inclination between the
post-kick inner and outer binaries is sufficiently high.

To investigate how the eccentricity and inclination distributions of stellar
triples translate into those of BH triples that survive all three SNs, we carry
out numerical calculations that we describe below. Since the orbital parameters
change due to SN explosion, we use superscripts ``0" and ``k" to denote the
pre-kick and post-kick orbital parameters. Our procedure is as follows:

\begin{itemize}

\item We begin with wide hierarchical stellar triples. The recent \textit{Gaia} data has revealed a significant population of wide binaries
in the galactic field \citep[e.g.,][]{Gaia}. One of the components could be a binary (i.e., inner binary$~+~$tertiary star). For our fiducial example, we consider a triple system consisting of $m_1^0=m_3^0=33.3M_\odot$ and $m_2^0=22.2M_\odot$.
To explore different hierarchical structures, we choose $a_\IN^0=(10, 30, 50,
100)\;\mathrm{AU}$ and $a_\OUT^0=(3, 5, 8, 10)\times10^3\;\mathrm{AU}$, respectively.
The level of the hierarchy, i.e., $a_\IN^0/a_\OUT^0$, therefore ranges from $0.003$ to $0.033$.
We assume the eccentricity of the inner binary, $e_\IN^0$, is randomly distributed in the range of $[0, 1]$,
as suggested in \citet{hwang2022} for binaries with semimajor axis $\lesssim
100\;\mathrm{AU}$.
For the outer binary eccentricity, $e_\OUT^0$, we sample about $\sim10^3$ values from the highly superthermal distribution given by \citet{xu2023wide}.
Later, we will also draw $e_\OUT^0$ values from the uniform and thermal distributions for comparison.
The mutual inclinations of the two orbits, $I_\OUT^0$, are less certain. On one hand, observed triple systems tend to exhibit a diverse range of mutual inclination angles \citep{Bork16}.
On the other hand, the medium where stars form is suggested to be not only turbulent but also magnetized
(e.g., \citealt{Crut10,Humc19,Xudyn20}).
Strong magnetic fields play an essential role in the formation of massive binary stars.
The magnetic braking effect can efficiently remove sufficient angular momentum during the gravitational contraction of matter
and enables the formation of a very close high-mass binary
(e.g., \citealt{Lund18,Hara21}).
As the efficiency of magnetic braking depends on the magnetic field orientation relative to the the rotation axis of the contracting matter
\citep{Henn09,Joos12,Liz13,Tsuka18}, it is likely that the orientation of
binary orbital angular momentum is regulated by the magnetic field and is not random.
Due to these uncertainties, we assume two types of distributions for the initial mutual inclination angles, i.e.,
$\cos I_\OUT^0$ is drawn from a uniform distribution either
over $[-1,1]$ (``isotropic'') or over $[-1, -\sqrt{3/5}]\cup[\sqrt{3/5}, 1]$
(``less-inclined''). In the latter case, the stellar triples considered do not
experience ZLK oscillations. Finally, the arguments of periapse ($\omega$) and
longitudes of ascending node ($\Omega$) for the two orbits are drawn from a
uniform distribution over $[0, 2\pi)$.

Throughout this procedure, we reject and regenerate any systems that are dynamically unstable. We use the condition given by \citet{mardling2001tidal}:
\be\label{eq:stability}
\frac{a_\OUT}{a_\IN}>2.8\bigg(1+\frac{m_3}{m_{12}}\bigg)^{2/5}\frac{(1+e_\OUT)^{2/5}}
{(1-e_\OUT)^{6/5}}\bigg(1-\frac{0.3 I_\OUT}{180^\circ}\bigg).
\ee
Note that the stability of multiple systems has been studied extensively, including improved characterization at high mutual inclinations \EDIT{\citep{2013Georga, hayashi_stab1, hayashi_stab2}}.
More recent constraints can be found in \EDIT{\citep{grishin_stab, vynatheya_stab, 2022tory, lalande_stab}}, but our results based on Eq.~\eqref{eq:stability}
are not expected to change significantly.

\item Next, we allow all three stars to undergo supernovae explosion (SNe).
For simplicity, we assume that all BHs born in our system retain $90\%$ of their progenitor masses, losing $10\%$ due to neutrino emission during their SNe.
For our fiducial parameters above, this results in remnant masses of $m_1^k=m_3^k=30M_\odot$ and $m_2^k=20M_\odot$.
We assume that the first SN explosion occurs on the primary star $m_1$ in the inner binary, and the following SNs occur on $m_3$ and $m_2$ in sequence.
For each stable triple system,
we simulate a SN explosion for each mass component.
Every time a kick due to mass-loss happens, we draw the two orbital mean anomalies (orbital phases) from uniform distributions.
At the end of this procedure, the resulting semi-major axes, eccentricities, and orientations of the angular momenta of both orbits
are evaluated based on the methodology outlined in \citet{liu2021hierarchical},
and dynamically unstable systems according to Eq.~\eqref{eq:stability} are removed.

\item
Since the changes in the orbital parameters depend on the geometry of the triple system
(e.g. argument of periapse, longitude of the ascending node, and orbital inclination),
to cover all possibilities,
we repeat the previous step for each triple 100 times.
Therefore, the statistical features of the surviving systems can be characterized by the accumulated data.
\end{itemize}

One concern not addressed in the procedure above is that the inner binary may merge due to ZLK oscillations prior to becoming a BHB if the mutual inclination is sufficiently large. Large mutual inclinations are common in the isotropic $\cos I_\OUT^0$ distribution and can still be obtained infrequently for the less-inclined case due to mass-loss induced kicks. Here, we argue that this concern is unlikely to significantly affect our results.
The ZLK timescale is given by
\begin{align}\label{eq:LK timescale}
t_\zlk={}&\frac{1}{n_\IN}\bigg(\frac{m_{12}}{m_3}\bigg)\bigg(\frac{a_\OUT}{a_\IN}\bigg)^3(1-e_\OUT^2)^{3/2}\\
\simeq{}&10^6 \mathrm{yrs}\bigg(\frac{m_{12}}{50M_\odot}\bigg)^{1/2}\bigg(\frac{m_3}{30M_\odot}\bigg)^{-1}\nonumber\\
&\bigg(\frac{a_\IN}{100\;\mathrm{AU}}\bigg)^{-3/2}\bigg(\frac{a_\OUT}{3000\;\mathrm{AU}}\bigg)^3(1-e_\OUT^2)^{3/2}\nonumber,
\end{align}
where $n_\IN=\sqrt{(Gm_{12})/a_\IN^3}$ is the mean motion of the inner binary.
Because we are interested in modestly and significantly hierarchical triples,
the eccentricity of the inner binary is excited on the timescale of
$t_\zlk\propto(a_\OUT/a_\IN)^3\gtrsim10^6$yrs for the majority of systems studied here; this is comparable to or
greater than the lifetime of the massive stars.
Moreover, the $e_\IN-$excitation is sensitive to the inclination angle,
which may vary significantly during a series of SNs.
Only when $I_\OUT^0\backsimeq90^\circ$ can $e_\IN$ reach its maximum, and the system may produce a stellar merger event.
Therefore, based on the system setup, a tiny fraction of systems in our calculations is expected to generate stellar mergers prior to BHB formation.
More discussion can be found in Section~\ref{ssec:stellar_zlk}.

\subsection{Post-kick eccentricities \texorpdfstring{$e_\OUT^k$}{eoutk}}\label{sec 2 3}

\begin{figure}
    \centering
    \includegraphics[width=\columnwidth]{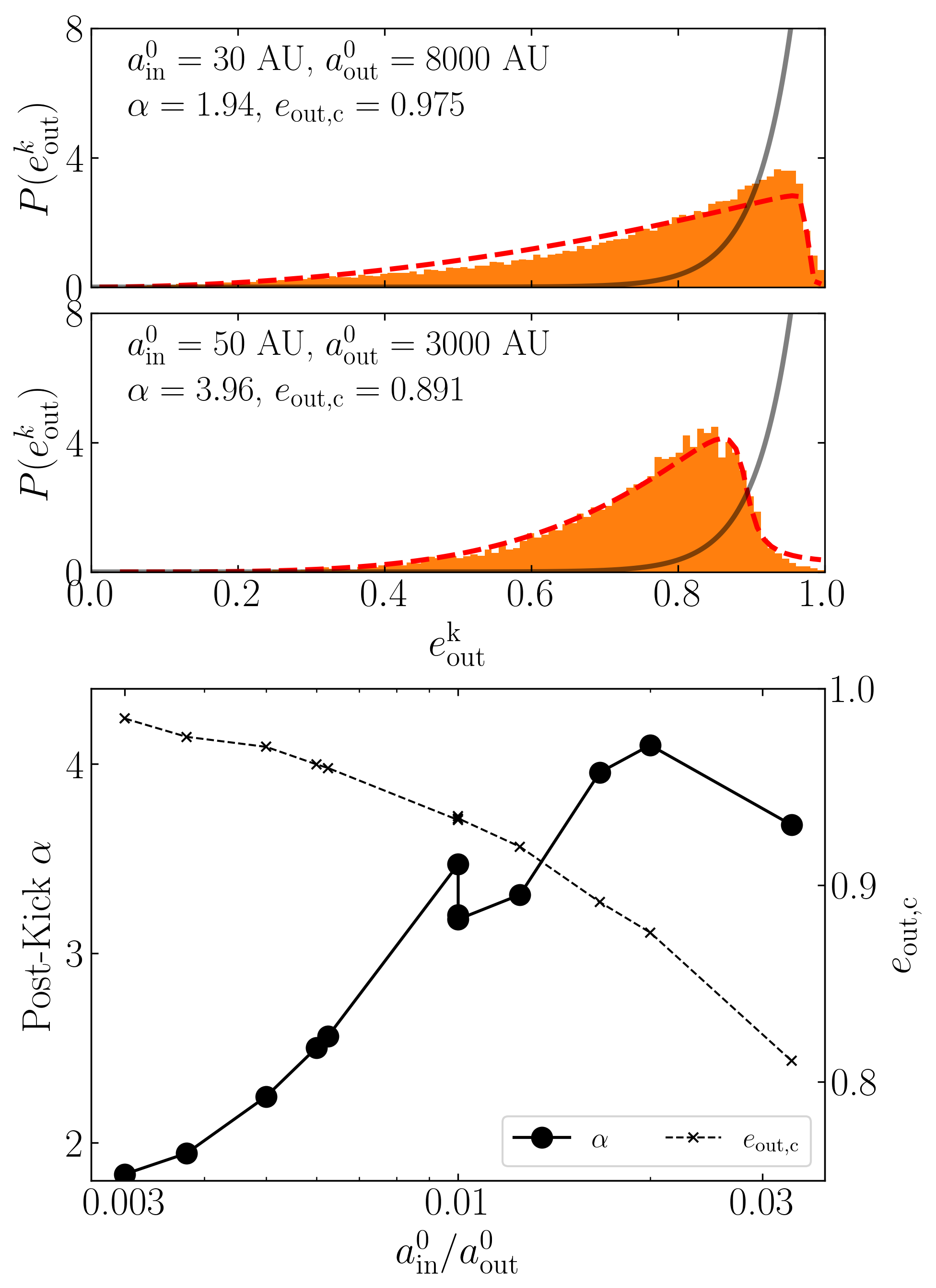}
    \caption{The top two panels show the distributions of the post-kick outer binary eccentricity ($e_{\rm out}^k$)
    as the orange filled histogram for the same orbital hierarchies as Fig.~\ref{fig:Post kick triples} (labelled). For comparison,
    the initial eccentricity distribution is shown as the grey line (as described by Eq.~\ref{eq:pe}),
    and the truncated power law fit (Eq.~\ref{eq:eout_f_dist}, with select fit parameters labelled) is shown as the dashed red line.
    The bottom panel shows the power law exponent $\alpha$ (filled circles)
    and the cutoff eccentricity $e_{\rm c}$ (crosses) as a function of the ratio $a_{\rm in}^0 / a_{\rm out}^0$,
    where $a_{\rm in}^0 \in [30, 50, 100]\;
    \mathrm{AU}$ and $a_{\rm out}^0 \in [3, 5, 8, 10]
    \times 10^3\;\mathrm{AU}$.
    Note that while equal ratios of $a_{\rm in}^0 / a_{\rm out}^0$ generally result in qualitatively similar distributions, the randomly sampled distributions can yield different best-fit values for $\alpha$ (see Section~\ref{sec 2 3} for additional details).}\label{fig:indicies}
\end{figure}

Fig.~\ref{fig:Post kick triples}
displays the orbital parameter distributions of surviving BH triples for two levels of orbital hierarchy.
The two panels shown here represent two qualitatively different hierarchies that we consider:
significantly hierarchical systems (right panel) and modestly hierarchical systems (left panel).
We see that the kicks induced by mass-loss can either increase or decrease $a_\OUT$,
while the peak is around the original value ($a_\OUT^0$).
The eccentricity of the orbit also changes when a SN occurs.
In general, we notice that the overall distribution of $e_\OUT^k$ is different for the significantly and modestly hierarchical cases.
As shown in the right panel, when $e_\OUT^k\lesssim 0.9$, the survived systems present a superthermal $e_\OUT$ distribution,
which follows a power-law distribution with a slope of about $2$ (see also the bottom panel of Fig.~\ref{fig:indicies}).
A deficit of large values of $e_\OUT^k$ ($e_\OUT^k\gtrsim 0.9$) is due to the requirement of dynamical stability of the initial stellar
triples---note that the horizontal dashed line in the central panel and right histogram, representing the dynamical stability condition given by Eq.~\ref{eq:stability}, marks where the eccentricity distribution begins to decrease.
When $a_\IN^0/a_\OUT^0$ becomes larger and we approach the modestly hierarchical regime,
the peak of $e_\OUT^k$ distribution shifts to $e_\OUT^k\simeq0.8$ (as shown in the left panel of Fig.~\ref{fig:Post kick triples}).
Compared to the significantly hierarchical case, the eccentricity distribution is steeper for $e_\OUT^k\lesssim 0.8$
(i.e., for functional form $p(e_\OUT^k) \propto (e_\OUT^k)^{\alpha}$, $\alpha \gtrsim 3$, indicating a highly superthermal distribution) and
more systems with $e_\OUT^k\gtrsim 0.8$ are removed due to the pre-kick stability sampling.
It is worth noting that the distribution of $e_\IN^k$ for the inner BHB are almost unchanged,
remaining approximately uniform over $(0, 1)$.
This is in agreement with the result of \citet{hills1983effects},
which shows that mass loss by $\delta m$ in a binary with total mass $M$ only affects the eccentricity distribution for $e \lesssim \delta m / M$, where $\delta m / M = 0.05$ for each SN in the inner binary.

To better understand and describe the effect of the three SNe on the outer orbit, we seek a quantitative description of the $e_\OUT^k$ distribution. First, we compare in Fig.~\ref{fig:indicies} the distributions of the initial eccentricity $e_\OUT^0$ and the post-SNe eccentricity $e_\OUT^k$.
In accordance with the existing literature on superthermal eccentricity distributions \citep[e.g.][]{hwang2022},
we wish to fit the distribution of $e_{\rm out}^k$ to a power law.
However, as is evident from the upper panels of Fig.~\ref{fig:indicies}, the final distribution deviates significantly from power-law-like behavior
at higher eccentricities due to the aforementioned orbital stability constraints.
As such, we instead fit the distribution of $e_{\rm out}^k$ to the following functional form:
\begin{equation}
    f(e_{\rm out}^k) \propto (e_{\rm out}^k)^{\alpha}
        \left[\frac{1}{2} - \frac{2}{\pi}\arctan
            \left(\frac{e_{\rm out}^k - e_{\rm out, c}}{\Delta e}\right)\right].\label{eq:eout_f_dist}
\end{equation}
This models a power law distribution with index $\alpha$ and a stability-imposed cutoff at $e_{\rm out, c}$ with width $\Delta e$.
These fits are displayed as the red dashed lines in the upper panels of Fig.~\ref{fig:indicies}.
While the fits do not agree completely with the distributions obtained from simulations,
they adequately capture the large range of power-law-like growth of the $e_{\rm out}^k$ distribution at smaller eccentricities.

Of greatest interest is the power law index $\alpha$ in Eq.~\eqref{eq:eout_f_dist},
which is generally thought to be $\approx 1.3$ for wide binaries in \textit{Gaia} \citep{hwang2022}.
However, in our data, the values of $\alpha$ are significantly more superthermal, as shown in the bottom panel of Fig.~\ref{fig:indicies}.
This is not altogether too surprising, since the initial $e_{\rm out}^0$ is almost entirely in excess of $\sim 0.8$
(upper panels of Fig.~\ref{fig:indicies}).
As such, we conclude that the post-kick outer eccentricity distribution is still strongly affected by the stellar outer eccentricity distribution.
However, the $e_{\rm out}^k$ distribution is comparatively less sensitive to the initial $e_{\rm out}^0$ distribution for $a_{\rm in}^0/a_{\rm out}^0 \lesssim 0.002$ (as can be seen from the top panel of Fig.~\ref{fig:Survival fraction}).

Note that the recovered values of $\alpha$ for the less-hierarchical systems $a_{\rm in}^0 \gtrsim 0.01 a_{\rm out}^0$ exhibit mild differences for similar ratios of $a_{\rm in}^0 / a_{\rm out}^0$.
This is because $\alpha$ is a poorly constrained parameter for the less-hierarchical systems, for which the weaker mass-loss-induced kicks do not populate the smaller ranges of $e_{\rm out}^k$ as efficiently as the more violent kicks in more hierarchical systems.
As such, the power law index $\alpha$, which is sensitive to the behavior of $e_{\rm out}^k$ down to small values, can fluctuate substantially depending on the specific realization of the random sampling.
Nevertheless, while the $\alpha$ values present some mild spread at these milder hierarchies, they are sufficiently well constrained to be $\gtrsim 3$ and can be compared to distributions in the literature.

\subsection{Dependence on the initial \texorpdfstring{$e_\OUT^0$}{eout0} distribution}\label{sec 2 4}

\begin{figure*}
\centering
\begin{tabular}{cccc}
\includegraphics[width=17.5cm]{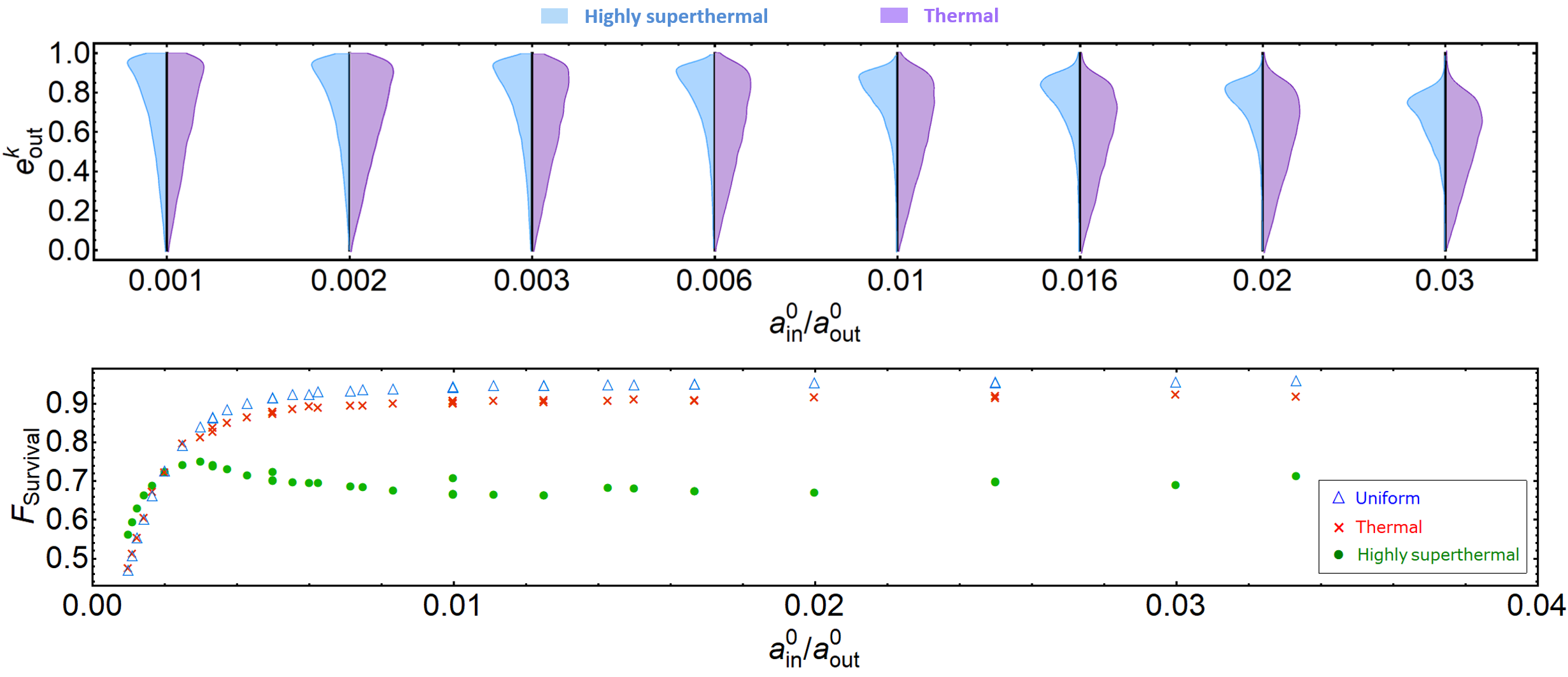}
\end{tabular}
\caption{Upper panel: the probability distributions of $e_\OUT^k$ for different values of $a_\IN^0/a_\OUT^0$.
We present two sets of initial $e_\OUT^0$ samples: thermal eccentricity distribution (purple) and highly superthermal distributions (blue).
Bottom panel: survival fraction of the triples that have experienced three SN explosions
as a function of hierarchical levels. The results are from the systems initialized with uniform eccentricity distribution,
thermal distribution and superthermal distribution, respectively (as labeled).
}\label{fig:Survival fraction}
\end{figure*}

In Section~\ref{sec 2 2}, we assumed that the initial $e_\OUT^0$ follows a highly superthermal distribution.
We now explore the impact of different types of $e_\OUT^0$ distributions on $e_\OUT^k$.
We repeat the analysis described in Section~\ref{sec 2 2} but with the initial outer eccentricities drawn from uniform and thermal ($P(e_\OUT^0) \propto e_\OUT^0$) distributions.
Our results are shown in Fig.~\ref{fig:Survival fraction}.

The upper panel shows the post-SNs $e_\OUT^k$ distribution as a function of $a_\IN^0/a_\OUT^0$.
The distributions shown in blue represent the results with the initial $e_\OUT^0$ sampled by Eq. (\ref{eq:pe}) as before.
Now, we also include the final $e_\OUT^k$ distribution from systems initialized with a thermal $e_\OUT^0$ distribution for each $a_\IN^0/a_\OUT^0$.
For the most extreme hierarchy $a_\IN^0/a_\OUT^0=10^{-3}$, we find that the two distributions present qualitatively similar profiles.
This is easily understood: the orbital velocities of the outer binary are relatively low and
the orbital parameters are significantly modified by the three SNe, erasing any memory of the initial $e_\OUT^0$ distribution.
The similary holds even for uniform $e_\OUT^0$ distributions (see more examples in Appendix~\ref{Appendix A}).
However, as the system becomes less hierarchical, the $e_\OUT^k$ distributions obtained for initially thermal and highly superthermal $e_\OUT^0$ distributions deviate increasingly.
It is worth noting that in the case of a highly superthermal $e_\OUT^0$ distribution,
the lack of systems distributed at $e_\OUT^k\lesssim0.5$ is a result of the initial sampling (see upper panels of Fig.~\ref{fig:indicies}).

The lower panel of Fig.~\ref{fig:Survival fraction} illustrates the survival fraction for different hierarchical structures.
Here, $F_\mathrm{survival}$ is the mean value of runs $\sim 10^5$ ($\sim 10^3$ values of $e_{\rm out}^0$, each sampled with $100$ geometries)
and is evaluated by comparing the number of stable triples right before/after the three kicks due to mass loss.
We see that the fractions $F_\mathrm{survival}$ given by the uniform and thermal distributions of $e_\OUT^0$ present similar values as a function of $a_\IN^0/a_\OUT^0$: $F_\mathrm{survival}$
is mostly constant for modest hierarchies and decreases dramatically as $a_\IN^0/a_\OUT^0\lesssim0.01$.
This is because the survival fraction depends sensitively on $e_\OUT^0$ (see the examples shown in Appendix~\ref{Appendix A}).
For eccentric orbits, the orbital velocity at the apocenter is lower than the velocity at the pericenter, and so the mass-loss induced kicks are more impactful.
Since mass loss occurs randomly in orbital phase,
eccentric binaries have a higher risk of being disrupted by SNe compared to circular binaries.
However, systems with large $a_\IN^0/a_\OUT^0$ have a higher outer orbital velocity, which suppresses the effect of the mass-loss induced kicks. As such, triples with modest hierarchical levels, i.e., $a_\IN^0/a_\OUT^0\gtrsim0.01$, are likely to survive all the SNs.
Note that the fraction $F_\mathrm{survival}$ produced by the highly superthermal $e_\OUT^0$ distribution
becomes lower than those from the other two cases.
This is a result of the large number of systems with extreme $e_\OUT^0$, which easily become unstable after SNe.
Detailed comparisons about the orbital response to mass-loss induced kicks can be found in Appendix~\ref{Appendix A}.

\subsection{Post-kick inclinations \texorpdfstring{$I_\OUT^k$}{ioutk}}\label{sec 2 5}

\begin{figure}
    \centering
    \includegraphics[width=\columnwidth]{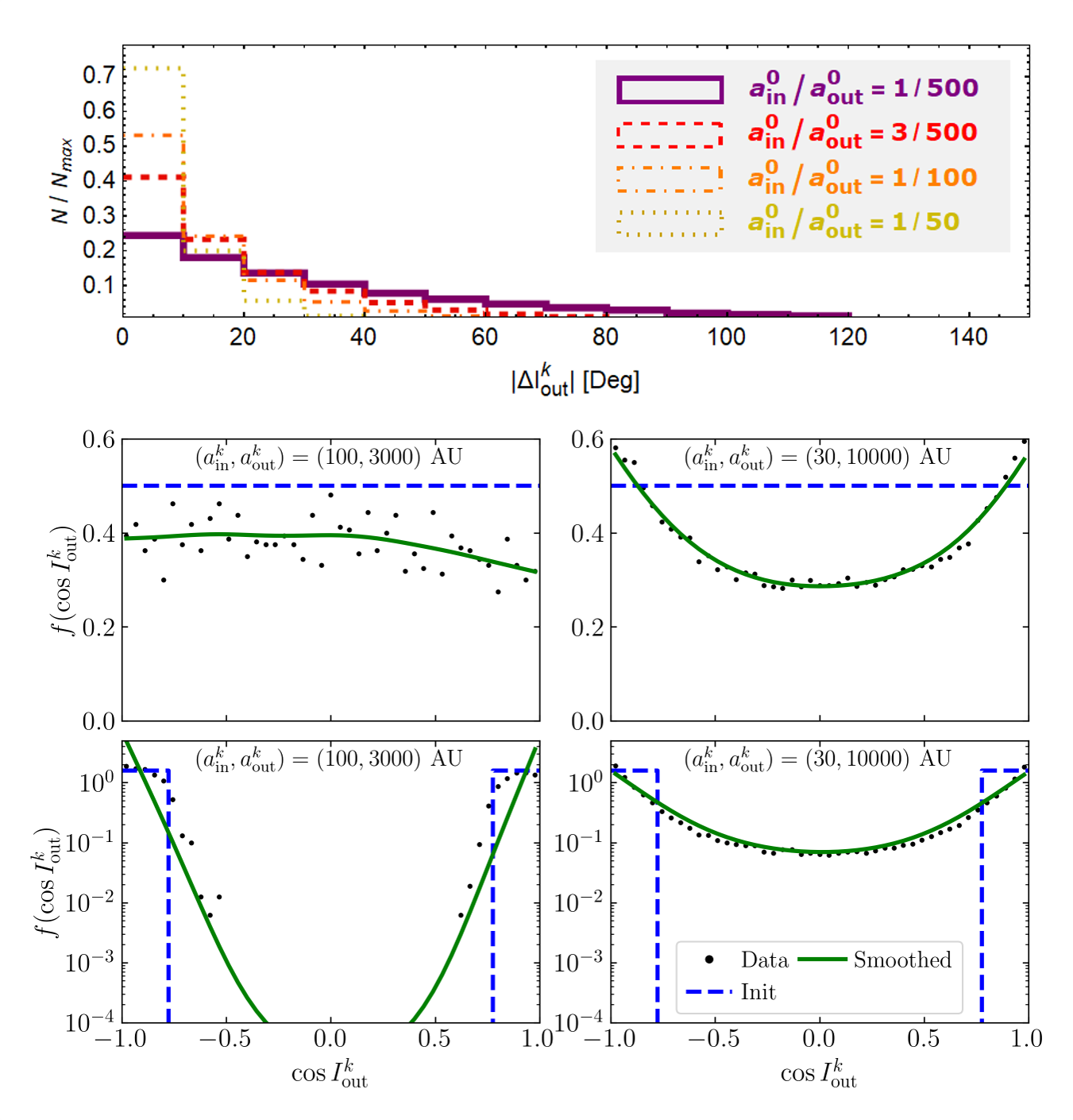}
    \caption{The top panel shows the probability distributions of
    the change in the mutual inclination angles $|\Delta I_\OUT^k|$, taking into account the different levels of hierarchy (color-coded; as labeled).
    The results are from the systems initialized with an isotropic distribution of $I_\OUT^0$.
    The middle pannels show the distribution of mutual inclinations between the inner and outer orbit ($I_{\rm out}^k$)
    for two different orbital hierarchies as the black dots, again assuming an isotropic distribution for $I_\OUT^0$ (blue dashed line).
    The smoothed fit to the $I_\OUT^k$ distribution is shown as the green line. Note that, while the initial (blue) distribution is normalized to $1$,
    the distribution of $I_{\rm out}^k$ (black dots) is normalized to the fraction of systems surviving all three SNe ($F_{\rm survival}$) and is $< 1$.
    The bottom pannels show the same but for the less-inclined $I_\OUT^0$ distribution (blue dashed line); note that the ZLK window can be filled in a modest amount by the mass-loss induced kicks.}\label{fig:Ihist2-unif}
\end{figure}

To examine how the mutual inclination of the inner and outer orbits can be varied due to kicks,
we study the changes in inclination angle $|\Delta I_\OUT^k|$ and the final angles $I_\OUT^k$ for a large range of $a_\IN^0/a_\OUT^0$.
The upper panel of Fig.~\ref{fig:Ihist2-unif}
shows the results of the triples with initial inclinations ranging from $0^\circ$ to $180^\circ$.
We observe a broad range of $|\Delta I_\OUT^k|$, indicating that the orientations of the angular momenta often change significantly.
Comparing the range of $|\Delta I_\OUT^k|$ for each $a_\IN^0/a_\OUT^0$, we note that more hierarchical triples result in larger changes in the angular momentum orientations.

In the middle panels of Fig.~\ref{fig:Ihist2-unif}, we show the post-kick distributions of $I_\OUT^k$ for an initially isotropic distribution of $I_\OUT^0$ [$P(\cos I_{\rm out}^0)$ is uniform].
It is counterintuitive that, even when the mutual inclination between the inner and outer binary orbits is initially isotropically distributed, the inclination distribution after all three supernovae can be noticeably non-isotropic.
This is most apparent when $a_{\rm in} / a_{\rm out}$ is smaller, such as in the middle-right panel of Fig.~\ref{fig:Ihist2-unif}.
This pattern is persistent for a wide range of parameters as can be found in Appendix~\ref{app:Idists}.

This behavior can be understood by studying the effect of the mass-loss induced kick on the outer binary---there is no mass-loss induced kick to the inner orbit.
First, the direction of the kick on the outer orbit is always in the plane of the inner orbit.
Moreover, only the component of this effective kick \textit{normal} to the outer orbit can change the mutual orbital inclination.
As such, if the kick magnitude is $v_k$, the maximum change in the mutual orbital inclination must scale as
\begin{equation}
    |\Delta I_{\rm out}^k|_{\max}
        \propto v_k\sin I_{\rm out}^0.\label{eq:Ioutmax}
\end{equation}
The actual change in mutual inclination depends on the orbital phase at the time of the supernova explosion, and the distribution of $I_\OUT^k$ is approximately symmetric about zero.
According to Eq.~\eqref{eq:Ioutmax}, if the two binaries are nearly coplanar, corresponding to $I_{\rm out}^0 \sim 0^\circ$ or $\sim 180^\circ$,
we see that the change in mutual inclination $\Delta I_{\rm out}$ is correspondingly small.
On the other hand, when the initial misalignment is large ($I_{\rm out}^0 \sim 90^\circ$),
the final misalignment angle $I_{\rm out}^k$ can vary over a much larger range.
As such, large mutual inclinations become less common after an SNe in the inner binary, and accordingly coplanar inclinations must become more common.
This is indeed the effect observed in the middle-right panel of Fig.~\ref{fig:Ihist2-unif}
(note that the integral of the green line is normalized to the fraction of systems that survive all three SNe and is thus $< 1$).

In the bottom panels of Fig.~\ref{fig:Ihist2-unif}, we also show the post-kick inclination distribution for systems that are not within the ZLK window during their stellar phase.
It can be seen that the ZLK window can be filled somewhat efficiently significantly hierarchical systems, where the larger mass-loss induced kicks can more easily reorient the outer binary angular momentum.

\section{Merging Black Hole Binaries}\label{s:popsynth}\label{sec 3}

Next, we study the orbital dynamics of the BH triples formed via the process
described in Section~\ref{sec 2}.
Most notably, we will compare the BH merger efficiency for the uniform and thermal distributions of $e_{\rm out}^k$ commonly used in the literature to the superthermal distributions of $e_{\rm out}^k$ obtained above.
In the interest of generality, we do not
directly integrate the outcomes from the previous section, and we instead study
the merger rates for various system parameters and distributions of orbital
elements. This allows us to isolate the effect of specific parameters on the
observed properties of merging BH triples. Our procedure follows that of \citet{su2021mass}, and we briefly describe it here.

We use the double-averaged secular equations of motion at the octupole level of
approximation as given by \citet{LML15}. This holds as long as the outer orbital
period satisfies\EDIT{
\begin{equation}
    P_{\rm out} \lesssim t_{\rm ZLK} (1 - e_{\max}^2)^{1/2},\label{eq:dblavg}
\end{equation}}
with $t_{\rm ZLK}$ given by Eq.~\eqref{eq:LK timescale}. Most of our systems lie
in the regime where the double-averaged equations are valid
(Eq.~\ref{eq:dblavg}) save for the most eccentric outer binaries. Strictly
speaking, these require more accurate integration techniques such as
single-averaged equations of motion or direct N-body integration
\citep[see][]{Antonini_2012, Antonini_2014, luo2016, lei2018, 2018grishin,
bin_misc5, LL19, Hamers, mangipudi2022}. However, the use of more approximate
equations of motion is not generally expected to affect merger rates
\citep{bin_misc5, su2021mass}, so we use the double-averaged equations for all
systems for simplicity. Additional long-term eccentricity oscillations can occur
when the period ratios of the two orbits is not so extreme \citep{luo2016,
lei2018, tremaine2023brown}, and can be incorporated into the double-averaged
Hamiltonian, but this effect is most relevant in the test-particle limit of ZLK
and is subdominant for our comparable-mass system. We also include general
relativistic apsidal precession of the inner binary, a first order
post-Newtonian (1PN) effect, and the emission of GWs, a 2.5PN effect. \EDIT{We integrate our equations using the \texttt{solve\_ivp} function of the \texttt{scipy} Python library, using the \texttt{LSODA} integrator using relative and absolute tolerances of $10^{-11}$.}

\subsection{Parameter Selection}\label{sec:params}\label{sec 3 1}

For the BH triple merger studies here, we consider the same BH masses as above
($m_1^k = m_3^k = 30M_\odot$ and $m_2^k = 20M_\odot$), and we consider three
values of $a_{\rm in}^k \in (30, 50, 100)\;\mathrm{AU}$ with four values of $a_{\rm out}^k
= (3, 5, 8, 10) \times 10^3\;\mathrm{AU}$.
For each pair of $a_{\rm in}^k$ and $a_{\rm out}^k$, we resample the remaining
parameters as follows: $e_{\rm in}$ is drawn from a uniform distribution,
$e_{\rm out}^k$ is drawn from a thermal distribution (again subject to the
stability constraint given by Eq.~\ref{eq:stability}; we consider other
$e_\OUT^k$ distributions below), $\cos I_{\rm out}^k$ is drawn from a uniform
distribution over $[-\sqrt{3/5}, \sqrt{3/5}]$ in the ZLK-active regime, and
finally the remaining orbital parameters ($\omega_{\rm in}$, $\omega_{\rm out}$,
$\Omega_{\rm in}$, $\Omega_{\rm out}$) are drawn randomly in the interval $[0,
2\pi)$.
A total of $2 \times 10^4$ systems are integrated for each of the $12$ orbital hierarchies of $(a_{\rm in}^k, a_{\rm out}^k)$.
If the system merges\footnote{\EDIT{We define merger to be $a_{\rm in} < 1\;\mathrm{AU}$, which is typically reached at very large eccentricity such that the inner binary subsequently merges rapidly within a negligible time \citep{su2020spin}.}} within a Hubble time ($10$ Gyr), its merger time $T_{\rm m}$ is recorded, and otherwise, it is considered unable to merge.

\begin{figure}
    \centering
    \includegraphics[width=\columnwidth]{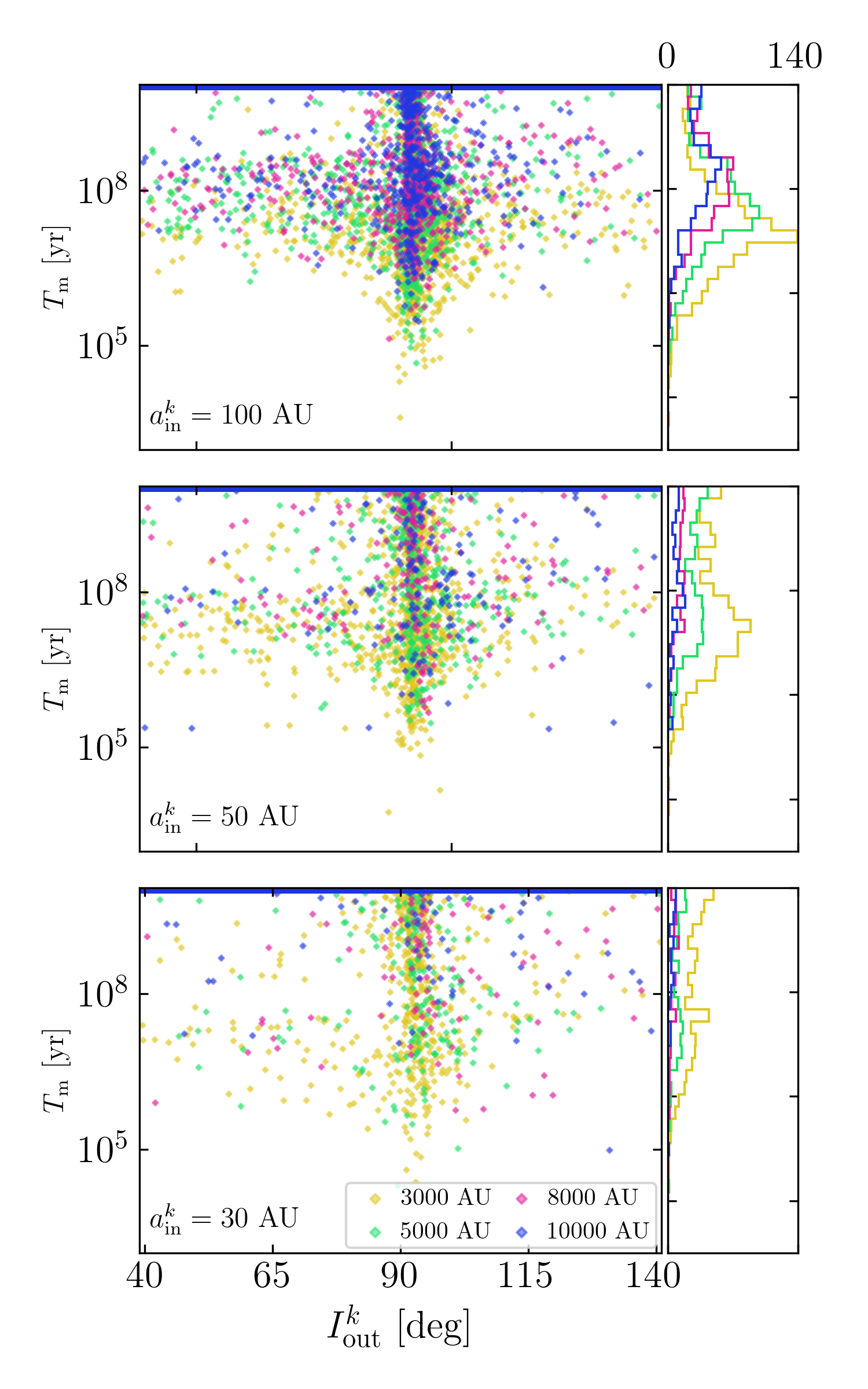}
    \caption{The left column plots show the merger times $T_{\rm m}$ as a
    function of the initial mutual inclination $I_{\rm out}^k$ between the inner
    and outer binaries where the inner eccentricity is uniformly sampled over
    $[0, 1)$, 
    and the outer eccentricity is sampled from a thermal distribution.
    Different colored dots denote different outer $a_{\rm out}^k$ (legend) and
    different panels denote different inner semimajor axes $a_{\rm in}^k$. For
    the remaining parameters, see Section~\ref{sec:params}. The right hand
    panels show the histograms of successful mergers. A total of $2 \times 10^4$
    cases are run.}\label{fig:tmerge_0}
\end{figure}

In Fig.~\ref{fig:tmerge_0}, we display the merger times obtained with this procedure for each combination of $a_{\rm in}^k$ and $a_{\rm out}^k$.
We comment on a few of the immediately apparent trends.
First, for each of the individual parameter combinations, it can be seen that most merging systems have $I_{\rm out}^k \sim 90^\circ$; such systems attain the most extreme eccentricities during ZLK cycles and thus emit the most gravitational wave radiation, leading to more mergers.
As for the few merging systems far from $I_{\rm out}^k \simeq 90^\circ$, these systems are unable to merge via the quadrupole-order ZLK effect alone; they instead undergo chaotic evolution due to the octupole-order ZLK effect, during which a fraction of systems will attain sufficiently large eccentricities to merge.
The characteristic merger time of such systems is the octupole-order ZLK timescale \citep{antognini2015timescales, su2021mass}, which corresponds to the peaks at $T_{\rm m} \sim 10\;\mathrm{Myr}$ in each of the histograms (right panels).
In the bottom-most panel for $a_\IN^k = 30\;\mathrm{AU}$, the peak is least prominent; this is because the octupole-order effect is smallest (see Eq.~\ref{eq:eps_oct}).

Next, when comparing the $T_{\rm m}$ merger plots across different systems, it is apparent that systems merge more efficiently for smaller $a_{\rm out}$ and larger $a_{\rm in}$.
The primary cause of this effect is the scaling of ZLK effect given by Eq.~\eqref{eq:LK timescale}, where shorter $t_{\rm ZLK}$ leads to more mergers.
Additionally, systems with larger $\epsilon_{\rm oct}$, given by Eq.~\eqref{eq:eps_oct}, are also more likely to merge.

\subsection{Merger Fraction}\label{ssec:merge_frac}\label{sec 3 2}

\begin{figure}
    \centering
    \includegraphics[width=\columnwidth]{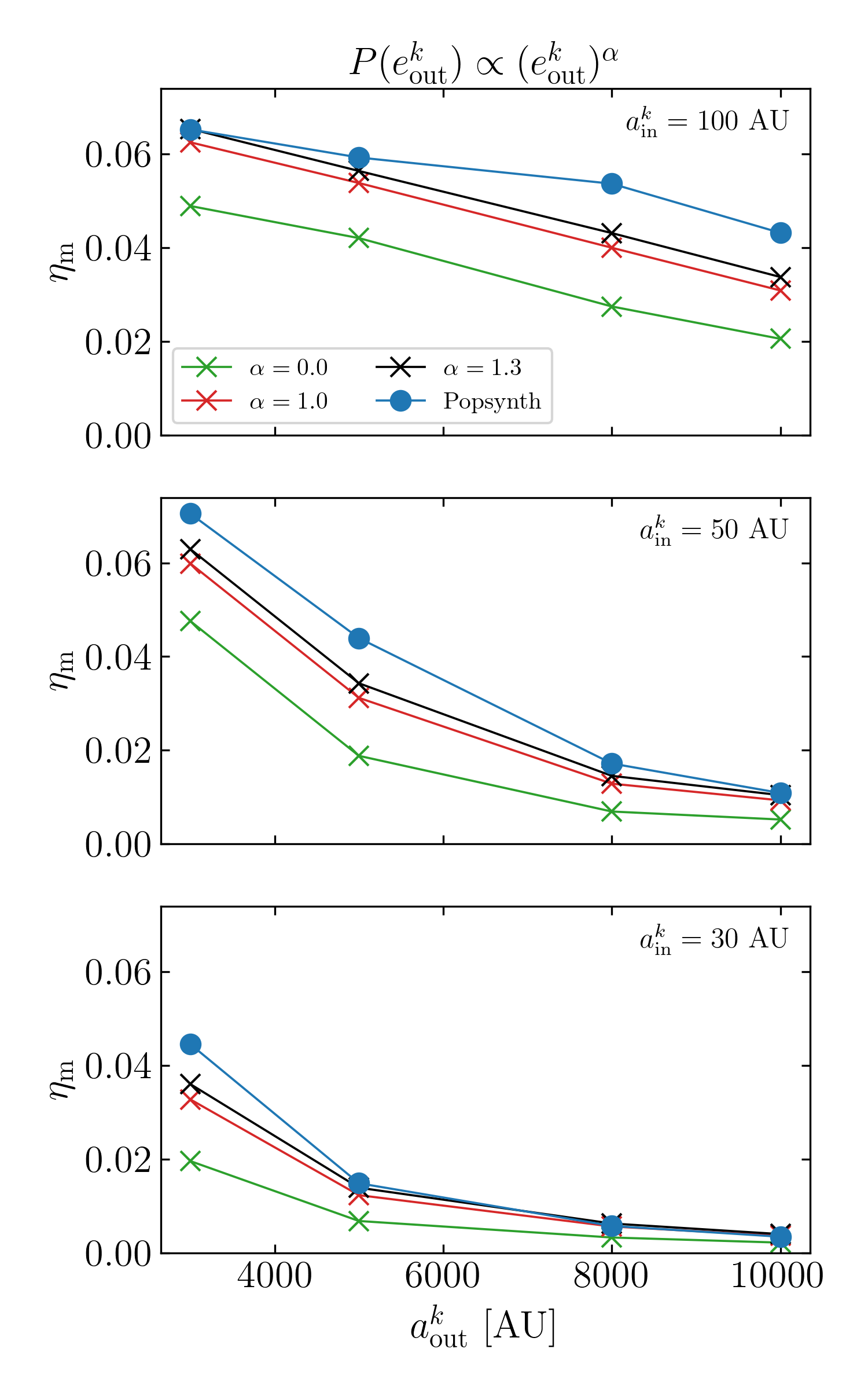}
    \caption{Merger fractions $\eta_{\rm m}$ (Eq.~\ref{eq:eta_merge}) for all orbital hierarchies considered in this paper. The different colored crosses correspond to different distributions of $e_{\rm out}^k$ (using Eq.~\ref{eq:eta_merge2}), the outer eccentricity upon formation of the black hole triple. Note that the colored crosses illustrate the merger fractions for power-law distributions $P(e_{\rm out}^k) \propto (e_{\rm out}^k)^\alpha$, while the blue circles are taken from the distributions obtained via the population synthesis described in Section~\ref{sec 2 3}.}\label{fig:etamerge_0}
\end{figure}

With these integrations, we can then compute the merger fraction $\eta_{\rm m}$ as the probability that a BH triple with a given $a_{\rm in}^k$ and $a_{\rm out}^k$ with a random $\cos I_{\rm out}^k \in [-1, 1]$ merges within $10$ Gyr. To be precise, we befine the merger fraction
\begin{equation}
    \eta_{\rm m} \equiv
    \frac{\sqrt{3/5}}{N}
    \sum_{i=1}^N
    \mathbf{1}_{\rm merge, i},\label{eq:eta_merge}
\end{equation}
where $\mathbf{1}_{\rm merge, i}$ denotes whether the $i$th integration merged; the factor of $\sqrt{3/5}$ accounts for the fact that the $N$ integrations are only drawn from within the ZLK window, and not the full $[-1,1]$ range.
This probability is shown in the red lines in Fig.~\ref{fig:etamerge_0}. We can see that $\eta_{\rm m}$ decreases with decreasing $a_{\rm in}^k$ and with increasing $a_{\rm out}^k$, in agreement with the discussion above in Section~\ref{sec:params}.

Note that different eccentricity distributions can be considered by re-weighting the probability that each integration contributes to $\eta_{\rm m}$ \citep[e.g.][]{su2021mass}: since the numerical integrations draw $e_{\rm out}^k$ from a thermal distribution, the merger fraction for some other distribution $P(e_{\rm out}^k)$ can be computed via
\begin{equation}
    \eta_{\rm m} \equiv
    \frac{\sqrt{3/5}}{N}
    \sum_{i=1}^N
    \frac{P(e_{\rm out}^k)}{e_{\rm out}^k}
    \mathbf{1}_{\rm merge, i},\label{eq:eta_merge2}
\end{equation}
a simple generalization of Eq.~\eqref{eq:eta_merge}.
Via this reweighting procedure, we can compute the merger probability for the
cases where $e_{\rm out}^k$ is uniformly distributed and where it is distributed
according to the mildly superthermal distribution found for wide stellar
binaries in \emph{Gaia} ($P(e_\OUT^k) \propto (e_\OUT^k)^{1.3}$;
\citealp{hwang2022}). The merger fractions for the uniform and mildly
superthermal $e_{\rm out}^k$ distributions are displayed as the green and black
lines respectively in Fig.~\ref{fig:etamerge_0}. We find that eccentricity
distributions that favor larger $e_{\rm out}^k$ result in enhanced merger
fractions due to the enhanced octupole-order effects (Eq.~\ref{eq:eps_oct}),
which is in agreement with the results of \citet{su2021mass}.

Note that the highly superthermal eccentricity distributions found in Section~\ref{sec 2} generally favor large $e_{\rm out}^k$ even more strongly than the mildly superthermal distribution above.
To study the impact of such highly superthermal $e_{\rm out}^k$ distributions found in Section~\ref{sec 2} on the BH merger rate, we can also compute $\eta_{\rm m}$ using the distributions of $e_{\rm out}^k$ obtained from Section~\ref{sec 2} for each $a_{\rm in}$ and $a_{\rm out}$ pair\footnote{Strictly speaking this procedure is not accurate, since the inner and outer semi-major axes change due to the SNe mass loss.
However, using the distributions of $a_{\rm in}^k$ and $a_{\rm out}^k$ would mix the effects of the orbital hierarchy on the BH phase and the effect of the three SNe. Therefore, for simplicity, we use the distributions of $e_{\rm out}^k$ for specific values of $a_{\rm in}^0$ and $a_{\rm out, 0}$.}.
In other words, we evaluate Eq.~\eqref{eq:eta_merge2} using $P(e_\OUT^k)$ from the procedures of Section~\ref{sec 2}, where some examples of $P(e_{\rm out}^k)$ are shown as the orange histograms in Fig.~\ref{fig:indicies}.
These results are displayed as the blue dots in Fig.~\ref{fig:etamerge_0}. We find that the highly superthermal distribution further enhances the merger rates but by only a modest amount.

\subsection{Effect of Orbital Evolution during Stellar Phase}\label{ssec:stellar_zlk}

Until now, we have directly taken the results of the post-kick orbital parameters and applied ZLK dynamics to them.
However, ZLK dynamics also occur before all three SNe are complete.
The detailed evolution including ZLK oscillations and stellar evolution in conjunction is very complex and uncertain.
While we argue in Section~\ref{sec 2 2} that the BH triples considered in this work form faster than the ZLK timescale, we can further assess the impact of stellar-phase ZLK-induced mergers by comparing this optimistic assumption to a conservative assumption where all ZLK-active stellar triples merge before forming BHs.
This corresponds to using the ``less-inclined'' $I_\OUT^0$ distribution defined above, where $\cos I_\OUT^0$ is drawn over $[-1, -\sqrt{3/5}]\cup[\sqrt{3/5}, 1]$.

In order to study the inclination dependence of BH mergers, we seek to incorporate the BH inclination distribution $f(I_\OUT^k)$, where we choose the convention where $f(I_\OUT^k)$ is normalized to the survival probability of a stellar triple after all three SNe, i.e.\ $F_{\rm survival}$ (Fig.~\ref{fig:Survival fraction}).
Then, we define the merger branching ratio
\begin{equation}
    \Gamma_{\rm m} \equiv
    \frac{\sqrt{3/5}}{N}
    \sum_{i=1}^N
    \frac{P(e_{\rm out}^k)}{e_{\rm out}^k}
    f(I_{\rm out}^k)
    \mathbf{1}_{\rm merge, i},\label{eq:eta_merge_Iout}
\end{equation}
This expression simply weights the contribution of each stellar triple to the aggregate merger probability by the probability density of its eccentricity and orbital misalignment.

\begin{figure}
    \centering
    \includegraphics[width=\columnwidth]{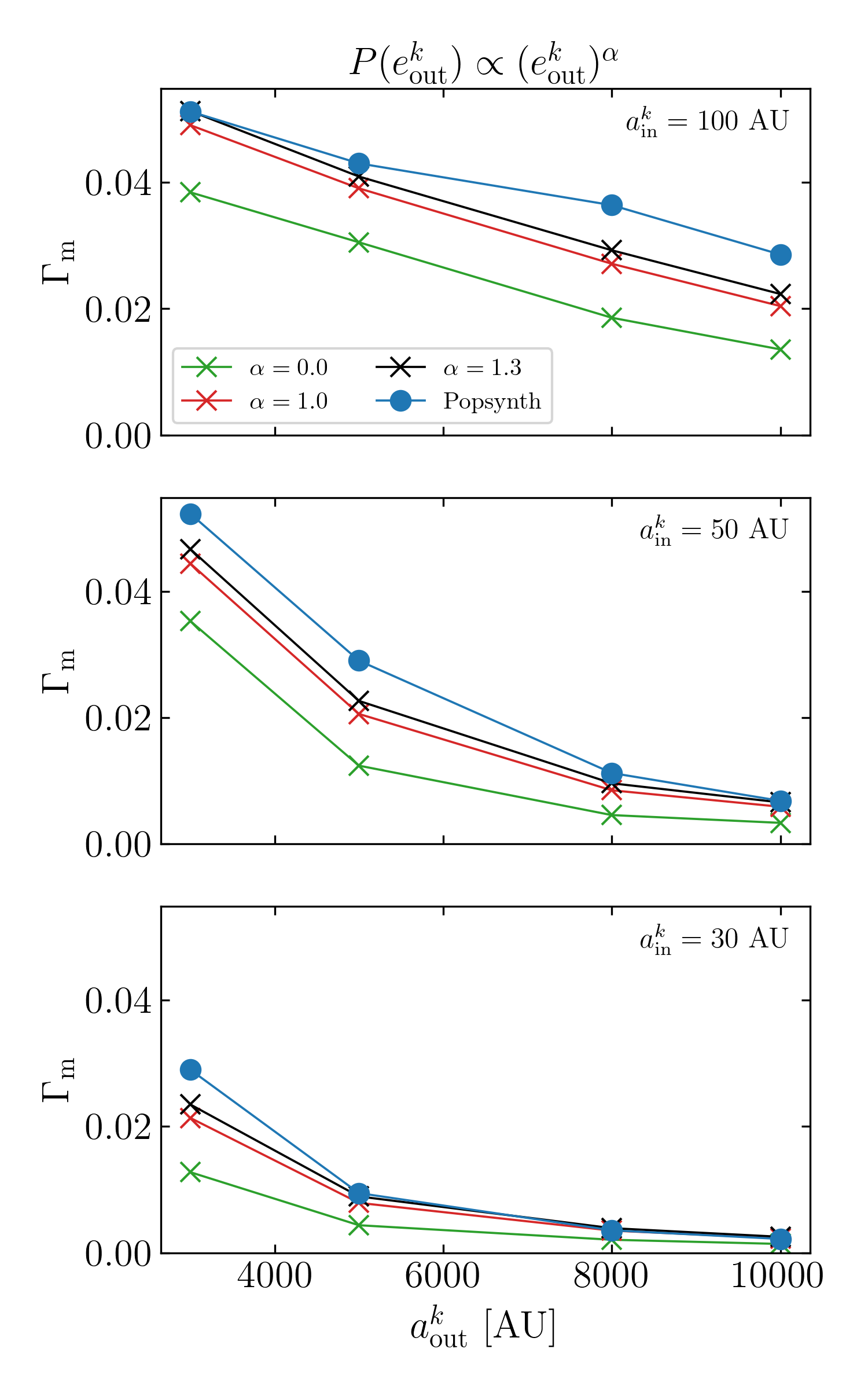}
    \caption{An extension of Fig.~\ref{fig:etamerge_0} but accounting for
    probability of reaching a given $I_{\rm out}^k$ from an initially
    isotropically distributed $I_{\rm out}^0$ (denoted $\Gamma_{\rm m}$ as given
    by Eq.~\ref{eq:eta_merge_Iout}).}\label{fig:etamerge_Iconv_0}
\end{figure}

We begin by considering the optimistic case, where $I_\OUT^0$ is isotropically distributed, and no stellar ZLK-induced mergers occur.
The merger branching ratios for this case are shown in Fig.~\ref{fig:etamerge_Iconv_0}.
The values of $\Gamma_{\rm m}$ are slightly suppressed from Fig.~\ref{fig:etamerge_0} by a factor of $F_{\rm survival}$, but the overall trends are not significantly changed.
However, note that the trends do differ slightly, since $\Gamma_{\rm m}$ is not proportional to $\eta_{\rm m}$: recall from Section~\ref{sec 2 5} that an isotropic distribution of $I_{\rm out}^0$ does not always yield an isotropic distribution of $I_{\rm out}^k$.

\begin{figure}
    \centering
    \includegraphics[width=\columnwidth]{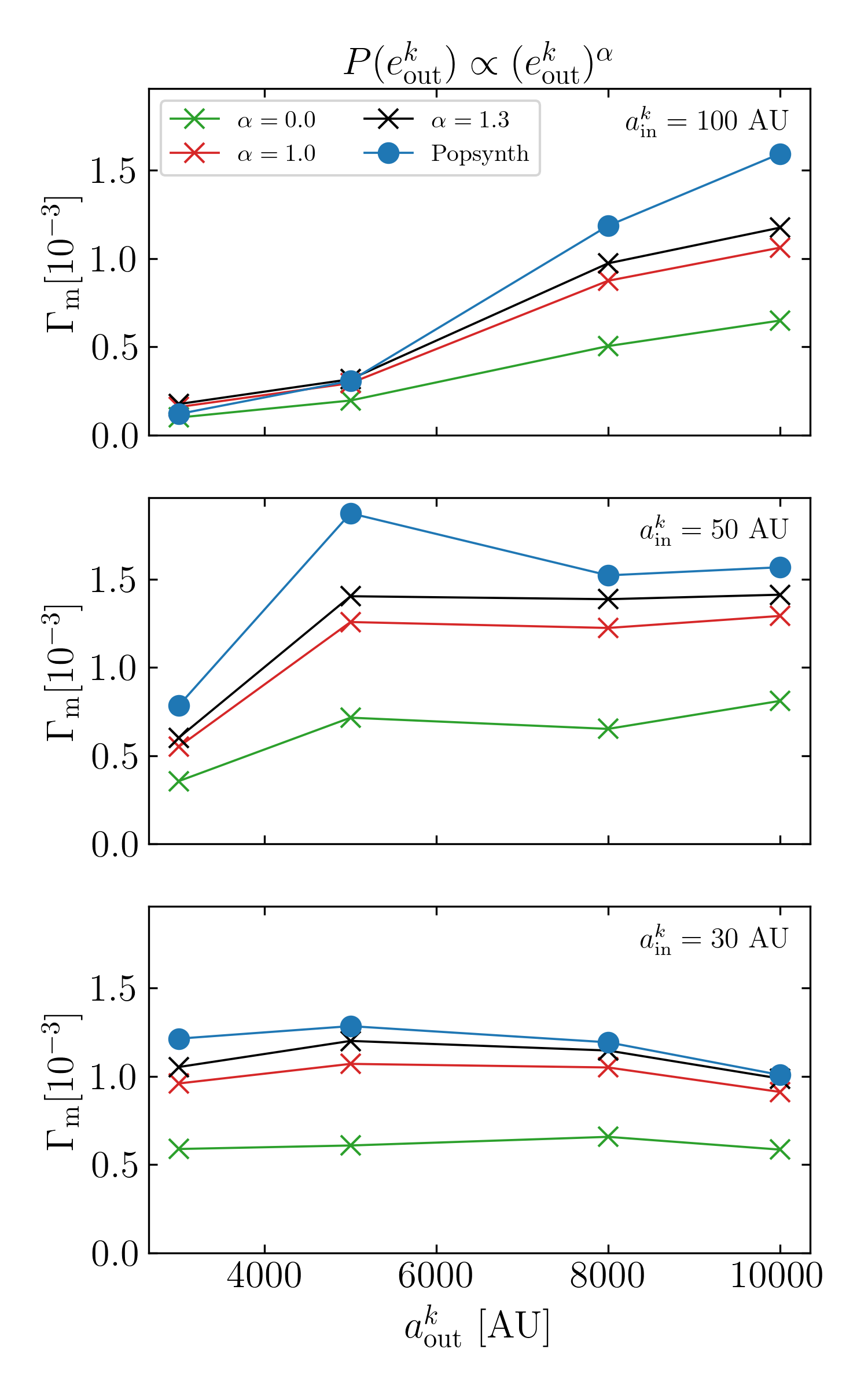}
    \caption{Same as Fig.~\ref{fig:etamerge_Iconv_0} but using only systems that
    reach a given $I_{\rm out}^k$ from an initial $I_{\rm out}^0$ not in the
    ZLK-active window $\cos I_{\rm out}^0 \in [-\sqrt{3/5},
    \sqrt{3/5}]$.}\label{fig:etamerge_Iconv_lessinc}
\end{figure}

We next consider the less-inclined $f(I_{\rm out}^0)$ distribution, corresponding to the conservative assumption where all stellar triples in the ZLK window merge. The resulting merger branching ratios $\Gamma_{\rm m}$ for these triples are shown in Fig.~\ref{fig:etamerge_Iconv_lessinc}.
There are two notable differences compared to the isotropic-$I_{\rm out}^{\rm 0}$ case. First, the typical branching ratio $\Gamma_{\rm m}$ is smaller by one to two orders of magnitude. This is because the merging systems predominantly have $I_{\rm out}^k$ close to $90^\circ$ (Fig.~\ref{fig:tmerge_0}), and such systems are much rarer when there are no primordial $I_{\rm out}^0$ in the ZLK window (see the green line in the bottom panels of Fig.~\ref{fig:Ihist2-unif}).
Second, note that $\Gamma_{\rm m}$ now does not vary monotonically with $a_{\rm out}^k$, and sometimes exhibits an increasing trend with larger $a_{\rm out}^k$ (top panel).
This is because of the effect discussed in Section~\ref{sec 2 5}: systems with larger $a_{\rm out}^k$ experience larger changes in $I_{\rm out}$ and are thus able to fill in the initially-empty ZLK window more efficiently than systems with smaller $a_{\rm out}^k$.
This combats the effect where larger $a_{\rm out}^k$ lead to weaker ZLK oscillations (as discussed in Section~\ref{ssec:merge_frac} and as is responsible for the decreasing trend in Fig.~\ref{fig:etamerge_Iconv_0}), and the competition between these two effects results in the parabolic shape of $\Gamma_{\rm m}$ in the bottom two panels of Fig.~\ref{fig:etamerge_Iconv_lessinc}.

\section{Summary and Discussion}\label{sec 4}

In this paper, we have re-examined the dynamical formation of merging BH binaries induced by a tertiary companion via the von Zeipel-Lidov-Kozai (ZLK)
effect, focusing on the effect of the outer eccentricity distribution.
In particular, we study the impact of beginning with a stellar triple where the outer orbit's eccentricity follows the extreme eccentricity distribution that has been
observed and physically justified for wide stellar binaries \citep{hwang2022, xu2023wide}.
In our scenario, we follow the process where each component in a wide stellar triple undergoes a SN explosion, and the entire system becomes a BH triple.
Mass-loss induced kicks (also called Blaauw kicks) can change the configuration of the hierarchical triples, or even break up the system.
We explore the BH triple systems that survive all three SNe.
We further use the post-kick parameters to study the
merger fraction in the channel of ZLK-induced merger.
The results of our calculations can be summarized as follows:

(i) The extreme eccentricity distributions of the initial stellar triples can be
modified by both mass-loss induced kicks and the three-body dynamical stability
criterion after each SN\@. The effect of the kick scales with the hierarchical
level of the triples. In general, for systems with $a_\IN/a_\OUT \gtrsim 0.005$,
the kick is not as important, and the fraction of outer binaries with extremely
high eccentricity is removed primarily due to stability. However, for triples
with $a_\IN/a_\OUT\lesssim 0.005$, the effect of mass-loss induced kick cannot
be ignored. The eccentricity varies dramatically and the final eccentricity
distributions lose most memory of their initial shapes. Quantitatively, this can
be described by similar power-law indicies for the final eccentricity
distributions when fit to a truncated power law of the form
Eq.~\eqref{eq:eout_f_dist}.

(ii) The post-kick mutual inclination, the angles between the two angular momenta of two orbits,
can vary significantly due to the mass-loss induced kicks.
The change of the inclination ($|\Delta I_\OUT^k|$) exhibits a wide range distribution ($\lesssim 120^\circ$),
depending on the hierarchical levels of the stellar triples, as shown in Fig.~\ref{fig:Ihist2-unif}.
Note that $|\Delta I_\OUT^k|$ is correlated to the change of the orbital eccentricity $|\Delta e_\OUT^k|$,
in which modestly/significantly hierarchical triples tend to produce mild/dramatic $|\Delta I_\OUT^k|$ and $|\Delta e_\OUT^k|$ respectively.
We also find that an initially isotropic inclination distribution can become noticeably non-isotropic after all SNs (see Fig.~\ref{fig:Ihist2-unif}).
This can be qualitatively understood by using Eq.~(\ref{eq:Ioutmax}).

(iii) We examine the effect of different outer binary eccentricity distributions on the final BH merger fractions.
By studying isotropically oriented inner and outer BH binaries.
a highly superthermal initial (stellar) $e_{\rm out}^0$ distribution enhances the merger fractions compared to a naive thermal distribution by a factor of $\lesssim 1.5$, depending on the specific system hierarchy
(see Fig.~\ref{fig:etamerge_0}).
Systems with larger $a_\IN^k$ and larger $a_\OUT^k$ (i.e.\ larger $\epsilon_{\rm oct}$, Eq.~\ref{eq:eps_oct}) merge more readily than their lower-$\epsilon_{\rm oct}$ counterparts.

(iv) For the dependence on the inclination angles,
in the case where the stellar triples have random orientations, the resulting BH merger fractions are comparable to the case where the BH triples themselves have random orientations (Fig.~\ref{fig:etamerge_Iconv_0}; as discussed above, these are not equivalent). However, when we restrict our attention to stellar triples that are not in the ZLK window, we find that strongly hierarchical systems (smaller $\epsilon_{\rm oct}$) are instead more susceptible to successful mergers---see Fig.~\ref{fig:etamerge_Iconv_lessinc}. In reality, some fraction of stellar triples in the ZLK window are likely to survive as BH triples, so the results of a complete population synthesis are likely bracketed by Figs.~\ref{fig:etamerge_Iconv_0} and~\ref{fig:etamerge_Iconv_lessinc}.

\subsection{Discussion}\label{ss:discussion}

In this work, we have made a number of simplifying assumptions to isolate the
effect of different outer eccentricity distributions on the merger dynamics of
the tertiary-induced merger channel. We discuss these assumptions and their
relation to the broader astrophysical literature here.

Most notably, we have neglected the effect of BH natal kicks in Section~\ref{sec
2}. These kicks are imparted to the BH upon its formation and are both poorly
understood and poorly observationally constrained \citep[e.g.][]{bray2018kick,
mandel2020kick}. There is some expectation that these are likely to be small for
the high-mass BHs in our fiducial scenario, which form via direct collapse (see
\citealp{Giacobbo2020} and references therein). Observationally, the best constraints
on the natal kicks are inferred from three binaries all consisting of a $\sim 10M_\odot$ BH and a stellar-
mass companion, and their conclusions differ: the
O-type star-BH binary system VFTS 243 is consistent with small natal kicks,
where estimates range from $4\;\mathrm{km/s}$ \citep{vignagomez2023natal} to $<
72\;\mathrm{km/s}$ (\citealp{banagiri2023}; consistent with a Blaauw kick), and the K giant-BH binary
system V404 Cygni is thought to have experienced a natal kick below $\sim 5\;\mathrm{km/s}$ to avoid
unbinding an observed tertiary companion at a separation of $\sim 3500\;\mathrm{AU}$
\citep[notably similar to the systems we consider;][]{burdge2024v404}. On
the other hand, the low-mass x-ray binary MAXI J1305-704 is thought to have
experienced a large natal kick $> 70\;\mathrm{km/s}$ due to its large peculiar
velocity \citep{kimball2023kick}. If BH natal kicks are larger than the
Keplerian orbital velocity of the outer binary ($v_{\rm kep} \sim 10$ km/s),
they may result in dissociation of the outer binary, decreasing the overall
merger fraction\footnote{For reference, under the common prescription where BH
natal kicks are scaled to neutron-star kicks by the ratio of their masses
[$v_{\rm k, bh} \sim (M_{\rm ns} / M_{\rm bh}) v_{\rm k, ns}$ where $v_{\rm k,
ns} \sim 200$--$300$ km/s and $M_{\rm ns} \sim 1.4M_\odot$], the resulting BH
natal kicks for our $M_{\rm bh} = 30M_\odot$ are $\approx 10$--$15$ km/s
\citep[e.g.][]{repetto2012}, though neutron star kick velocities are also
uncertain \citep[e.g.][]{coleman2022kicks}.}. Even if they are only of order
$v_{\rm kep}$, they will still result in substantial changes to the outer binary
orbital elements, largely mitigating the impact of the assumed outer
eccentricity distribution $e_{\rm out}^0$. Since many works already study the
tertiary-induced merger channel with natal kicks
\citep[e.g.][]{liu2021hierarchical}, we maximize the effect of outer eccentricity distribution on the merger dynamics by neglecting these
kicks.

We also have not considered the effect of different component masses or mass
ratios, though the effect of the latter is studied in other works
\citep[e.g.][]{su2021mass, martinez2021mass}. We assume a uniform mass loss
fraction during supernovae, which is an oversimplification but again eases interpretation of the effect of the outer eccentricity distribution; see e.g.\
\citet{spera2015mass, spera2017very} for more realistic BH remnant properties.
In particular, if the objects retain less mass during SNe (smaller BH masses), the mass-loss induced kicks will be stronger, resulting in a lower $F_{\rm survival}$ in general. In addition, the transition between ``less hierarchical'' and ``more hierarchical'' (where less hierarchical systems retain their primordial $e_{\rm out}^0$ distributions and more hierarchical systems are shaped by the SNe to have closer-to-thermal $e_{\rm out}^k$ distributions) will occur at larger ratios of $a_{\rm in}^0 / a_{\rm out}^0$.
We've neglected the stellar triple phase, arguing that it is short compared to the ZLK timescales in our fiducial systems, but \citet{kummer2023} find a nonzero (albeit small) fraction of systems experiencing mass transfer on the main sequence phase for our fiducial system hierarchies.
Detailed work beyond the scope of this paper would be required to understand whether, for particular system hierarchies, mass transfer primarily depletes: highly-misaligned systems, which attain larger eccentricities (and would reduce the final BH merger rate), or mildly-misaligned systems, which have a shorter ZLK period (and would have little effect on the final BH merger rate).
Our integrations
use the double-averaged equations of motion for the dynamical evolution in
Section~\ref{s:popsynth}, but this is not expected to affect the overall merger
fractions significantly \citep{bin_misc5}. Finally, we have not considered the
survival rate of wide stellar triples in their birth cluster environment.
Many of these changes would be required to make quantitative forecasts of merger fractions and rates
associated with the tertiary-induced merger channel, and we defer these to future work.


\section*{Acknowledgements}

\EDIT{We thank the anonymous referee, whose comments and suggestions improved the quality of this work.}
YS is supported by a Lyman Spitzer, Jr. Postdoctoral Fellowship at
Princeton University.
BL gratefully acknowledges support from
the European Union's Horizon 2021 research and innovation programme under the Marie Sklodowska-Curie grant agreement No. 101065374.
\EDIT{We thank Evgeni Grishin for helpful comments.}

\section*{Data Availability}

The data underlying this article will be shared on reasonable request to the corresponding author.



\bibliography{main}{}

\begin{thebibliography}{}
\expandafter\ifx\csname natexlab\endcsname\relax\def\natexlab#1{#1}\fi
\providecommand{\url}[1]{\href{#1}{#1}}
\providecommand{\dodoi}[1]{doi:~\href{http://doi.org/#1}{\nolinkurl{#1}}}
\providecommand{\doeprint}[1]{\href{http://ascl.net/#1}{\nolinkurl{http://ascl.net/#1}}}
\providecommand{\doarXiv}[1]{\href{https://arxiv.org/abs/#1}{\nolinkurl{https://arxiv.org/abs/#1}}}

\bibitem[{Abbott {et~al.}(2023)Abbott, Abe, Acernese, Ackley, Adhicary,
  Adhikari, Adhikari, Adkins, Adya, Affeldt, Agarwal, Agathos, Aguiar, Aiello,
  Ain, Ajith, Akutsu, Albanesi, Alfaidi, Al-Jodah, Alléné, Allocca, Almualla,
  Altin, Amato, Amez-Droz, Amorosi, Anand, Ananyeva, Andersen, Anderson,
  Anderson, Andia, Ando, Andrade, Andres, Andrés-Carcasona, Andrić, Ansoldi,
  Antelis, Antier, Aoumi, Apostolatos, Appavuravther, Appert, Apple, Arai,
  Araya, Araya, Areeda, Arène, Aritomi, Arnaud, Arogeti, Aronson, Arun, Asada,
  Ashton, Aso, Assiduo, de~Souza~Melo, Aston, Astone, Aubin, AultONeal, Babak,
  Badalyan, Badaracco, Badger, Bae, Bagnasco, Bai, Baier, Baiotti, Baird,
  Bajpai, Baka, Ball, Ballardin, Ballmer, Baltus, Banagiri, Banerjee, Bankar,
  Baral, Barayoga, Barber, Barish, Barker, Barneo, Barone, Barr, Barsotti,
  Barsuglia, Barta, Barthelmy, Barton, Bartos, Basak, Basalaev, Bassiri, Basti,
  Bawaj, Bayley, Baylor, Bazzan, Bécsy, Bedakihale, Beirnaert, Bejger, Bell,
  Benedetto, Beniwal, Benoit, Bentley, Yaala, Bera, Berbel, Bergamin, Berger,
  Bernuzzi, Beroiz, Berry, Bersanetti, Bertolini, Betzwieser, Beveridge,
  Bevins, Bhandare, Bhandari, Bhardwaj, Bhatt, Bhattacharjee, Bhaumik, Bianchi,
  Bilenko, Bilicki, Billingsley, Bini, Birnholtz, Biscans, Bischi, Biscoveanu,
  Bisht, Biswas, Bitossi, Bizouard, Blackburn, Blair, Blair, Blair, Bobba,
  Bode, Boër, Bogaert, Boileau, Boldrini, Bolingbroke, Bonavena, Bondarescu,
  Bondu, Bonilla, Bonilla, Bonnand, Booker, Bork, Boschi, Bose, Bose,
  Bossilkov, Boudart, Bouffanais, Bozzi, Bradaschia, Brady, Braglia, Branch,
  Branchesi, Brau, Breschi, Briant, Brillet, Brinkmann, Brockill, Brooks,
  Brooks, Brown, Brunett, Bruno, Bruntz, Bryant, Bucci, Buchanan, Bulashenko,
  Bulik, Bulten, Buonanno, Burtnyk, Buscicchio, Buskulic, Buy, Byer, Davies,
  Cabras, Cabrita, Cadonati, Caesar, Cagnoli, Cahillane, Bustillo, Callaghan,
  Callister, Calloni, Camp, Canepa, Santoro, Cannavacciuolo, Cannon, Cao, Cao,
  Capistran, Capocasa, Capote, Carapella, Carbognani, Carlassara, Carlin,
  Carpinelli, Carter, Carullo, Diaz, Casentini, Castaldi, Castro-Lucas,
  Caudill, Cavaglià, Cavalieri, Cella, Cerdá-Durán, Cesarini, Chaibi,
  Chakalis, Subrahmanya, Champion, Chan, Chan, Chandra, Chang, Chang, Chanial,
  Chao, Chapman-Bird, Charlton, Charlton, Chassande-Mottin, Chastain,
  Chatterjee, Chatterjee, Chatterjee, Chaturvedi, Chaty, Chatziioannou, Chen,
  Chen, Chen, Chen, Chen, Chen, Chen, Chen, Cheng, Chessa, Cheung, Chia,
  Chiadini, Chiang, Chiang, Chiarini, Chiba, Chiba, Chierici, Chincarini,
  Chiofalo, Chiummo, Choudhary, Christensen, Chua, Chung, Ciani, Ciecielag,
  Cieślar, Cifaldi, Ciobanu, Ciolfi, Clara, Clark, Clarke, Clearwater, Clesse,
  Cleva, Coccia, Codazzo, Cohadon, Colleoni, Collette, Colombo, Colpi, Compton,
  Conti, Cooper, Corban, Corbitt, Cordero-Carrión, Corezzi, Cornish, Corsi,
  Cortese, Coschizza, Cottingham, Coughlin, Coulon, Countryman, Coupechoux,
  Cousins, Couvares, Coward, Cowart, Cowburn, Coyne, Coyne, Craig, Creighton,
  Creighton, Criswell, Crockett-Gray, Croquette, Crowder, Cudell, Cullen,
  Cumming, Cummings, Cuoco, Curyło, Dabadie, Canton, Dall’Osso, Dálya,
  D’Angelo, Danilishin, D’Antonio, Danzmann, Darroch, Darsow-Fromm,
  Dasgupta, Datrier, Datta, Dattilo, Dave, Davenport, Davier, Davis, Davis,
  Daw, Dax, DeBra, Deenadayalan, Degallaix, Laurentis, Deléglise, Favero,
  Lillo, Lillo, Dell’Aquila, Pozzo, Matteis, D’Emilio, Demos, Dent,
  Depasse, Pietri, Rosa, Rossi, DeSalvo, Simone, Dhurandhar, Diab, Diamond,
  Díaz, Didio, Dietrich, Fiore, Fronzo, Giorgio, Giovanni, Giovanni, Girolamo,
  Diksha, Lieto, Michele, Pace, Palma, Renzo, Divyajyoti, Dmitriev, Doctor,
  Dohmen, Doleva, Donahue, D’Onofrio, Donovan, Dooley, Dooney, Doravari,
  Dorosh, Drago, Driggers, Drori, Ducoin, Dunn, Dupletsa, Durante, D’Urso,
  Duverne, Dwyer, Eassa, Easter, Ebersold, Eckhardt, Eddolls, Edelman, Edo,
  Edy, Effler, Eichholz, Eisenmann, Eisenstein, Ejlli, Engelby, Engl, Errico,
  Essick, Estellés, Estevez, Etzel, Evans, Evans, Evans, Evstafyeva, Ewing,
  Fabrizi, Faedi, Fafone, Fair, Fairhurst, Fan, Fan, Farah, Farr, Farr,
  Fauchon-Jones, Favaro, Favata, Fays, Feicht, Fejer, Fenyvesi, Ferguson,
  Fernandez-Galiana, Ferrante, Ferreira, Fidecaro, Figura, Fiori, Fiori,
  Fishbach, Fisher, Fittipaldi, Fiumara, Flaminio, Fleischer, Fleming, Floden,
  Fong, Font, Fornal, Forsyth, Franke, Frasca, Frasconi, Freed, Frei, Freise,
  Freitas, Frey, Fritschel, Frolov, Fronzé, Fujimoto, Fukunaga, Fulda, Fyffe,
  Gabbard, Gabella, Gadre, Gaglani, Gair, Gais, Galaudage, Gallardo, Gamba,
  Ganapathy, Ganguly, Gao, Gaonkar, Garaventa, Garcia-Bellido, García-Núñez,
  García-Quirós, Gardner, Gargiulo, Garufi, Gasbarra, Gateley, Gayathri,
  Gemme, Gennai, George, Gerberding, Gergely, Ghonge, Ghosh, Ghosh, Ghosh,
  Ghosh, Ghosh, Giacoppo, Giaime, Giardina, Gibson, Gier, Giri, Gissi,
  Gkaitatzis, Glanzer, Gleckl, Glotin, Godfrey, Godwin, Goetz, Goetz, Golomb,
  Goncharov, González, Gosselin, Gouaty, Gould, Goyal, Grace, Grado, Graham,
  Granata, Granata, Gras, Grassia, Gray, Gray, Greco, Green, Green, Green,
  Green, Gretarsson, Gretarsson, Griffith, Griffiths, Griggs, Grignani,
  Grimaldi, Grote, Gruson, Guerra, Guetta, Guidi, Guimaraes, Gulati,
  Gulminelli, Gunny, Guo, Guo, Gupta, Gupta, Gupta, Gupta, Gupta, Gupta, Gurs,
  Gushima, Gustafson, Gutierrez, Guzman, Haegel, Hain, Haino, Halim, Hall,
  Hamilton, Hammond, Han, Haney, Hanks, Hanna, Hannam, Hannuksela, Hansen,
  Hanson, Harada, Harder, Haris, Harmark, Harms, Harry, Harry, Hartwig,
  Haskell, Haster, Hathaway, Haughian, Hayakawa, Hayama, Hayes, Healy,
  Heffernan, Heidmann, Heintze, Heinze, Heinzel, Heitmann, Hellman, Hello,
  Helmling-Cornell, Hemming, Hendry, Heng, Hennes, Hennig, Hennig, Henshaw,
  Vivanco, Heurs, Hewitt, Higginbotham, Hild, Hill, Himemoto, Hines, Hirata,
  Hirose, Ho, Hochheim, Hofman, Hohmann, Holcomb, Holland, Holley-Bockelmann,
  Hollows, Holmes, Holt, Holz, Hong, Hornung, Hoshino, Hough, Hourihane,
  Howell, Howell, Hoy, Hoyland, Hsieh, Hsieh, Hsiung, Hsu, Hu, Hu, Huang,
  Huang, Huang, Huang, Hübner, Huddart, Hughey, Hui, Hui, Husa, Huttner,
  Huxford, Huynh-Dinh, Hyland, Iakovlev, Iandolo, Idzkowski, Iess, Inayoshi,
  Inoue, Iorio, Iosif, Irwin, Isi, Ismail, Itoh, Iyer, JaberianHamedan,
  Jacqmin, Jacquet, Jadhav, Jadhav, Jain, Jain, James, Jan, Jani, Janiurek,
  Janquart, Janssens, Janthalur, Jaraba, Jaranowski, Jarov, Jasal, Jaume,
  Javed, Jenkins, Jenner, Jennings, Jia, Jiang, Liu, Jin, Johansmeyer, Johns,
  Johnson, Johnston, Johny, Jones, Jones, Jones, Jones, Jones, Joshi, Ju, Jung,
  Junker, Juste, Kajita, Kalaghatgi, Kalogera, Kamai, Kamiizumi, Kanda,
  Kandhasamy, Kang, Kanner, Kapadia, Kapasi, Karat, Karathanasis, Karki,
  Kasamatsu, Kas-danouche, Kashyap, Kasprzack, Kastaun, Kato, Katsanevas,
  Katsavounidis, Katsuren, Katzman, Kaur, Kawabe, Kawazoe, Kéfélian, Keitel,
  Kellard, Kelley-Derzon, Kennington, Key, Khadka, Khalili, Khan, Khanam,
  Khazanov, Khursheed, Kijbunchoo, Kim, Kim, Kim, Kim, Kim, Kim, Kim, Kim,
  Kimball, Kimura, Kinley-Hanlon, Kirchhoff, Kissel, Kiyota, Klimenko, Klinger,
  Knee, Knust, Kobayashi, Koch, Koehlenbeck, Koekoek, Kohri, Kokeyama, Koley,
  Koliadko, Kolitsidou, Kolstein, Kondrashov, Kong, Kontos, Korobko, Kossak,
  Kouvatsos, Kovalam, Koyama, Kozak, Kranzhoff, Kranzhoff, Kringel, Krishnendu,
  Królak, Kuehn, Kuijer, Kukihara, Kulkarni, Kumar, Kumar, Kumar, Kumar,
  Kumar, Kume, Kuns, Kuroyanagi, Kuwahara, Kwak, Lacaille, Lagabbe, Laghi,
  Lakkis, Lalande, Lalleman, Lamberts, Landry, Lane, Lang, Lange, Lantz, Rana,
  Rosa, Lartaux-Vollard, Lasky, Lawrence, Laxen, Lazzarini, Lazzaro, Leaci,
  Leavey, LeBohec, Lecoeuche, Lee, Lee, Lee, Lee, Lee, Lee, Lee, Legred,
  Lehmann, Lehner, Lemaître, Lenti, Leonardi, Leonova, Leroy, Letendre,
  Lethuillier, Levesque, Levin, Leyde, Li, Li, Li, Li, Lin, Lin, Lin, Lin, Lin,
  Lin, Lin, Lin, Linde, Linker, Littenberg, Liu, Liu, Llamas, Lo, Lo, London,
  Longo, Lopez, Portilla, Lorenzini, Loriette, Lormand, Losurdo, Lott, Lough,
  Loughlin, Lousto, Lovelace, Lowry, Lück, Lumaca, Lundgren, Lung, Lussier,
  Lynam, Ma, Ma, Ma’arif, Macas, MacInnis, Macleod, MacMillan, Macquet,
  Hernandez, Magazzù, Magee, Maggiore, Magnozzi, Mahesh, Mahesh, Maini,
  Majorana, Makarem, Maliakal, Malik, Man, Mandic, Mangano, Mannix, Mansell,
  Mansingh, Manske, Mantovani, Mapelli, Marchesoni, Pina, Marion, Márka,
  Márka, Markakis, Markosyan, Markowitz, Maros, Marquina, Marsat, Martelli,
  Martin, Martin, Martinez, Martinez, Martinez, Martinez, Martinovic, Martynov,
  Marx, Masalehdan, Mason, Masserot, Reid, Mastrodicasa, Mastrogiovanni,
  Mateu-Lucena, Matiushechkina, Matsunaga, Mavalvala, McCarthy, McClelland,
  McClincy, McCormick, McCuller, McGhee, McGinn, McIsaac, McIver, McLeod,
  McRae, McWilliams, Meacher, Mehmet, Mehta, Meijer, Melatos, Mendell,
  Menendez-Vazquez, Menoni, Mercer, Mereni, Merfeld, Merilh, Merritt,
  Merzougui, Messenger, Messick, Meyers, Meylahn, Mhaske, Miani, Miao,
  Michaloliakos, Michel, Michimura, Middleton, Mihaylov, Miller, Miller,
  Miller, Miller, Millhouse, Mills, Milotti, Minenkov, Mio, Mir,
  Miravet-Tenés, Mishra, Mishra, Mishra, Mistry, Mitchell, Mitra, Mitrofanov,
  Mitselmakher, Mittleman, Miyakawa, Miyoki, Mo, Modafferi, Moguel, Mohapatra,
  Mohite, Molina-Ruiz, Mondal, Mondin, Montani, Moore, Moragues, Moraru,
  Morawski, More, More, Moreno, Moreno, Morisaki, Moriwaki, Morras, Moscatello,
  Mours, Mow-Lowry, Mozzon, Muciaccia, Mukherjee, Mukherjee, Mukherjee,
  Mukherjee, Mukund, Mullavey, Munch, Muñiz, Murray, Murray-Dean, Muusse,
  Nadji, Nagar, Nagar, Nagarajan, Nakamura, Nakano, Nakano, Nakayama, Napolano,
  Nardecchia, Narikawa, Narola, Naticchioni, Nayak, Neil, Neilson, Nelson,
  Nelson, Nery, Nesseris, Neunzert, Ng, Ng, Nguyen, Nguyen, Nguyen, Nguyen,
  Quynh, Nichols, Nieradka, Nishino, Nishizawa, Nissanke, Nitoglia, Niu,
  Nocera, Norman, North, Novak, Siles, Nurbek, Nuttall, Oberling, O’Dell,
  Oelker, Oertel, Oganesyan, Oh, Oh, Oh, O’Hanlon, Ohashi, Ohashi, Ohkawa,
  Ohme, Ohta, Oliveira, Oliveri, Oohara, O’Reilly, Ormiston, Ormsby, Orselli,
  O’Shaughnessy, O’Shea, Oshima, Oshino, Ossokine, Osthelder, Ottaway,
  Overmier, Pace, Pagano, Page, Pai, Pai, Pal, Palashov, Pálfi, Palomba, Pan,
  Panda, Pang, Pannarale, Pant, Panther, Paoletti, Paoli, Paolone, Papalexakis,
  Pappas, Parisi, Park, Parker, Pascucci, Pasqualetti, Passaquieti, Passuello,
  Patel, Pathak, Patra, Patricelli, Patron, Paul, Payne, Pearce, Pedraza,
  Pedurand, Pegna, Pegoraro, Pele, Arellano, Penn, Perego, Pereira, Perez,
  Périgois, Perkins, Perreca, Perriès, Perry, Pesios, Petermann, Petrillo,
  Pfeiffer, Pham, Pham, Phukon, Phurailatpam, Piccinni, Pichot, Piendibene,
  Piergiovanni, Pierini, Pierra, Pierro, Pillant, Pillas, Pilo, Pinard,
  Pineda-Bosque, Pinto, Piotrzkowski, Piotrzkowski, Pirello, Pitkin, Placidi,
  Placidi, Planas, Plastino, Poggiani, Polini, Pompili, Pong, Ponrathnam,
  Porcelli, Portell, Porter, Posnansky, Poulton, Powell, Powell, Pracchia,
  Pradier, Prajapati, Prasai, Prasanna, Pratten, Principe, Prodi, Prokhorov,
  Prosposito, Prudenzi, Puecher, Pullin, Punturo, Puosi, Puppo, Pürrer, Qi,
  Quetschke, Quinonez, Quitzow-James, Raab, Raaijmakers, Radulesco, Raffai,
  Rail, Raja, Rajan, Ramirez, Ramirez, Ramos-Buades, Rana, Rana, Randel,
  Rangnekar, Rapagnani, Ray, Raymond, Raza, Razzano, Read, Regimbau, Rei, Reid,
  Reid, Reitze, Relton, Renzini, Rettegno, Revenu, Reza, Rezac, Rezaei, Ricci,
  Richards, Richardson, Rijal, Riles, Riley, Rinaldi, Robertson, Robertson,
  Robinet, Rocchi, Rodriguez, Rolland, Rollins, Romanelli, Romano, Romel,
  Romero, Romero-Shaw, Romie, Ronchini, Roocke, Rosa, Rosauer, Rose, Rosińska,
  Ross, Rossello, Roussel, Rowan, Rowlinson, Roy, Royzman, Rozza, Ruggi,
  Morales, Ruiz-Rocha, Ryan, Sachdev, Sadecki, Sadiq, Saffarieh, Saha, Saha,
  Saito, Sakai, Sakellariadou, Sako, Sakon, Salafia, Salces-Carcoba, Salconi,
  Saleem, Salemi, Sallé, Samajdar, Sanchez, Sanchez, Sanchez, Sanchis-Gual,
  Sanders, Sanuy, Saravanan, Sarin, Sasli, Sassi, Sassolas, Satari, Sauter,
  Savage, Savant, Sawada, Sawant, Sayah, Schaetzl, Scheel, Scherf, Scheuer,
  Schiworski, Schmidt, Schmidt, Schmitz, Schnabel, Schneewind, Schofield,
  Schönbeck, Schuler, Schulte, Schutz, Schwartz, Scott, Scott, Seetharamu,
  Seglar-Arroyo, Sekiguchi, Sellers, Sengupta, Sentenac, Seo, Sequino, Sergeev,
  Servignat, Setyawati, Shaffer, Shahriar, Shaikh, Shams, Shao, Sharma,
  Chaudhary, Shawhan, Shcheblanov, Sheela, Shen, Shepard, Sheridan, Shikano,
  Shikauchi, Shimizu, Shimode, Shinkai, Shoemaker, Shoemaker, ShyamSundar,
  Sider, Siegel, Sieniawska, Sigg, Silenzi, Singer, Singh, Singh, Singh,
  Singha, Sintes, Sipala, Skliris, Slagmolen, Slaven-Blair, Smetana, Smith,
  Smith, Smith, Soldateschi, Somala, Somiya, Soni, Soni, Sordini, Sorrentino,
  Sorrentino, Sotani, Soulard, Souradeep, Sowell, Spagnuolo, Spencer, Spera,
  Spinicelli, Srivastava, Srivastava, Stachie, Stachurski, Steer, Steinlechner,
  Steinlechner, Stergioulas, StPierre, Strang, Stratta, Strong, Strunk,
  Sturani, Stuver, Suchenek, Sudhagar, Sueltmann, Sugiyama, Suh, Sullivan,
  Summerscales, Sun, Sunil, Sur, Suresh, Sutton, Suzuki, Suzuki, Swinkels, Syx,
  Szczepańczyk, Szewczyk, Tacca, Tagoshi, Tait, Takahashi, Takahashi,
  Takamori, Takano, Takeda, Takeda, Talbot, Talbot, Tamaki, Tamanini, Tanabe,
  Tanaka, Tanaka, Tanasijczuk, Tanioka, Tanner, Tao, Tao, Tapia, Martín,
  Tarafder, Taranto, Taruya, Tasson, Teloi, Tenorio, Terhune, Terkowski,
  Themann, Thirugnanasambandam, Thomas, Thomas, Thomas, Thomas, Thompson,
  Thondapu, Thorne, Thrane, Tiwari, Tiwari, Tiwari, Toivonen, Tolley, Tomaru,
  Tomita, Tomura, Tonelli, Torres-Forné, Torrie, e~Melo, Tournefier,
  Trapananti, Travasso, Traylor, Trenado, Trevor, Tringali, Tripathee, Troiano,
  Trovato, Trozzo, Trudeau, Tsang, Tsang, Tse, Tso, Tsuchida, Tsukada, Tsutsui,
  Turbang, Turconi, Turski, Tuyenbayev, Ubach, Ubhi, Uchikata, Uchiyama, Udall,
  Uehara, Ueno, Unnikrishnan, Ushiba, Utina, Vahlbruch, Vaidya, Vajente,
  Vajpeyi, Valdes, Valentini, Vallero, Valsan, van Bakel, van Beuzekom, van
  Dael, van~den Brand, Broeck, Vander-Hyde, van~der Sluys, de~Walle, van
  Dongen, van Haevermaet, van Heijningen, Vanosky, van Putten, van Ranst, van
  Remortel, Vardaro, Vargas, Varma, Vasúth, Vecchio, Vedovato, Veitch, Veitch,
  Venneberg, Venugopalan, Verdier, Verkindt, Verma, Verma, Vermeulen, Veske,
  Vetrano, Viceré, Vidyant, Viets, Vijaykumar, Villa-Ortega, Vina, Vincent,
  Vinet, Viret, Virtuoso, Vitale, Vocca, Voigt, von Reis, von Wrangel, Vorvick,
  Vyatchanin, Wade, Wade, Wagner, Walet, Walker, Wallace, Wallace, Wang, Wang,
  Wang, Ward, Warner, Was, Washimi, Washington, Watada, Watarai, Watchi, Wayt,
  Weaver, Weaving, Webster, Weinert, Weinstein, Weiss, Weller, Weller,
  Wellmann, Wen, Weßels, Wette, Whelan, White, Whiting, Whittle, Wilk, Wilken,
  Willetts, Williams, Williams, Williamson, Willis, Willke, Wipf, Woan,
  Woehler, Wofford, Wong, Wong, Wong, Wright, Wu, Wu, Wu, Wysocki, Xiao, Xu,
  Yadav, Yamada, Yamamoto, Yamamoto, Yamamoto, Yamamoto, Yamamoto, Yamashita,
  Yamazaki, Yang, Yang, Yang, Yap, Yeeles, Yelikar, Yeung, Yokoyama, Yokozawa,
  Yoo, Yu, Yu, Yuzurihara, Zadrożny, Zannelli, Zanolin, Zeeshan, Zeidler,
  Zelenova, Zendri, Zevin, Zhang, Zhang, Zhang, Zhang, Zhang, Zhao, Zhao, Zhao,
  Zheng, Zhong, Zhou, Zhu, Zhu, Zimmerman, Zucker, Zweizig, (The LIGO
  Scientific~Collaboration, \& the KAGRA~Collaboration)}]{LIGO_O3b}
Abbott, R., Abe, H., Acernese, F., {et~al.} 2023, ApJ Supplement Series, 267,
  29, \dodoi{10.3847/1538-4365/acdc9f}

\bibitem[{Antognini(2015)}]{antognini2015timescales}
Antognini, J.~M. 2015, MNRAS, 452, 3610, \dodoi{10.1093/mnras/stv1552}

\bibitem[{Antonini {et~al.}(2014)Antonini, Murray, \& Mikkola}]{Antonini_2014}
Antonini, F., Murray, N., \& Mikkola, S. 2014, ApJ, 781, 45,
  \dodoi{10.1088/0004-637x/781/1/45}

\bibitem[{Antonini \& Perets(2012)}]{Antonini_2012}
Antonini, F., \& Perets, H.~B. 2012, ApJ, 757, 27,
  \dodoi{10.1088/0004-637x/757/1/27}

\bibitem[{Antonini {et~al.}(2017)Antonini, Toonen, \&
  Hamers}]{antonini2017binary}
Antonini, F., Toonen, S., \& Hamers, A.~S. 2017, ApJ, 841, 77,
  \dodoi{10.3847/1538-4357/aa6f5e}

\bibitem[{{Atallah} {et~al.}(2024){Atallah}, {Weatherford}, {Trani}, \&
  {Rasio}}]{atallah2024}
{Atallah}, D., {Weatherford}, N.~C., {Trani}, A.~A., \& {Rasio}, F. 2024, arXiv
  e-prints, arXiv:2402.12429, \dodoi{10.48550/arXiv.2402.12429}

\bibitem[{{Banagiri} {et~al.}(2023){Banagiri}, {Doctor}, {Kalogera}, {Kimball},
  \& {Andrews}}]{banagiri2023}
{Banagiri}, S., {Doctor}, Z., {Kalogera}, V., {Kimball}, C., \& {Andrews},
  J.~J. 2023, \apj, 959, 106, \dodoi{10.3847/1538-4357/ad0557}

\bibitem[{Banerjee {et~al.}(2010)Banerjee, Baumgardt, \&
  Kroupa}]{banerjee2010stellar}
Banerjee, S., Baumgardt, H., \& Kroupa, P. 2010, MNRAS, 402, 371,
  \dodoi{10.1111/j.1365-2966.2009.15880.x}

\bibitem[{Belczynski {et~al.}(2010)Belczynski, Dominik, Bulik, O'Shaughnessy,
  Fryer, \& Holz}]{belczynski2010effect}
Belczynski, K., Dominik, M., Bulik, T., {et~al.} 2010, ApJL, 715, L138,
  \dodoi{10.1088/2041-8205/715/2/L138}

\bibitem[{Belczynski {et~al.}(2016)Belczynski, Holz, Bulik, \&
  O'Shaughnessy}]{belczynski2016first}
Belczynski, K., Holz, D.~E., Bulik, T., \& O'Shaughnessy, R. 2016, Nature, 534,
  512

\bibitem[{{Blaauw}(1961)}]{blaauw1961}
{Blaauw}, A. 1961, \bain, 15, 265

\bibitem[{Blaes {et~al.}(2002)Blaes, Lee, \& Socrates}]{blaes2002kozai}
Blaes, O., Lee, M.~H., \& Socrates, A. 2002, ApJ, 578, 775

\bibitem[{{Borkovits} {et~al.}(2016){Borkovits}, {Hajdu}, {Sztakovics},
  {Rappaport}, {Levine}, {B{\'\i}r{\'o}}, \& {Klagyivik}}]{Bork16}
{Borkovits}, T., {Hajdu}, T., {Sztakovics}, J., {et~al.} 2016, \mnras, 455,
  4136, \dodoi{10.1093/mnras/stv2530}

\bibitem[{{Bray} \& {Eldridge}(2018)}]{bray2018kick}
{Bray}, J.~C., \& {Eldridge}, J.~J. 2018, \mnras, 480, 5657,
  \dodoi{10.1093/mnras/sty2230}

\bibitem[{{Burdge} {et~al.}(2024){Burdge}, {El-Badry}, {Kara}, {Canizares},
  {Chakrabarty}, {Frebel}, {Millholland}, {Rappaport}, {Simcoe}, \&
  {Vanderburg}}]{burdge2024v404}
{Burdge}, K.~B., {El-Badry}, K., {Kara}, E., {et~al.} 2024, arXiv e-prints,
  arXiv:2404.03719.
\newblock \doarXiv{2404.03719}

\bibitem[{{Chamandy} \& {Shukurov}(2020)}]{Cham20}
{Chamandy}, L., \& {Shukurov}, A. 2020, Galaxies, 8, 56,
  \dodoi{10.3390/galaxies8030056}

\bibitem[{{Coleman} \& {Burrows}(2022)}]{coleman2022kicks}
{Coleman}, M. S.~B., \& {Burrows}, A. 2022, \mnras, 517, 3938,
  \dodoi{10.1093/mnras/stac2573}

\bibitem[{{Crutcher} {et~al.}(2010){Crutcher}, {Wandelt}, {Heiles},
  {Falgarone}, \& {Troland}}]{Crut10}
{Crutcher}, R.~M., {Wandelt}, B., {Heiles}, C., {Falgarone}, E., \& {Troland},
  T.~H. 2010, \apj, 725, 466, \dodoi{10.1088/0004-637X/725/1/466}

\bibitem[{De~Mink \& Mandel(2016)}]{de2016chemically}
De~Mink, S., \& Mandel, I. 2016, MNRAS, 460, 3545

\bibitem[{Dominik {et~al.}(2012)Dominik, Belczynski, Fryer, Holz, Berti, Bulik,
  Mandel, \& O'Shaughnessy}]{dominik2012double}
Dominik, M., Belczynski, K., Fryer, C., {et~al.} 2012, ApJ, 759, 52

\bibitem[{Dominik {et~al.}(2013)Dominik, Belczynski, Fryer, Holz, Berti, Bulik,
  Mandel, \& O'Shaughnessy}]{dominik2013double}
---. 2013, ApJ, 779, 72

\bibitem[{Dominik {et~al.}(2015)Dominik, Berti, O'Shaughnessy, Mandel,
  Belczynski, Fryer, Holz, Bulik, \& Pannarale}]{dominik2015double}
Dominik, M., Berti, E., O'Shaughnessy, R., {et~al.} 2015, ApJ, 806, 263

\bibitem[{Downing {et~al.}(2010)Downing, Benacquista, Giersz, \&
  Spurzem}]{downing2010compact}
Downing, J., Benacquista, M., Giersz, M., \& Spurzem, R. 2010, MNRAS, 407, 1946

\bibitem[{Ford {et~al.}(2000)Ford, Kozinsky, \& Rasio}]{ford2000secular}
Ford, E.~B., Kozinsky, B., \& Rasio, F.~A. 2000, ApJ, 535, 385

\bibitem[{Fragione \& Kocsis(2019)}]{fragione2019}
Fragione, G., \& Kocsis, B. 2019, MNRAS, 486, 4781,
  \dodoi{10.1093/mnras/stz1175}

\bibitem[{Fragione \& Loeb(2019)}]{fragione2019loeb}
Fragione, G., \& Loeb, A. 2019, MNRAS, 486, 4443

\bibitem[{{Gaia Collaboration} {et~al.}(2023){Gaia Collaboration}, {Arenou},
  {Babusiaux}, {Barstow}, {Faigler}, {Jorissen}, {Kervella}, {Mazeh},
  {Mowlavi}, {Panuzzo}, {Sahlmann}, {Shahaf}, {Sozzetti}, {Bauchet},
  {Damerdji}, {Gavras}, {Giacobbe}, {Gosset}, {Halbwachs}, {Holl}, {Lattanzi},
  {Leclerc}, {Morel}, {Pourbaix}, {Re Fiorentin}, {Sadowski}, {S{\'e}gransan},
  {Siopis}, {Teyssier}, {Zwitter}, {Planquart}, {Brown}, {Vallenari}, {Prusti},
  {de Bruijne}, {Biermann}, {Creevey}, {Ducourant}, {Evans}, {Eyer}, {Guerra},
  {Hutton}, {Jordi}, {Klioner}, {Lammers}, {Lindegren}, {Luri}, {Mignard},
  {Panem}, {Randich}, {Sartoretti}, {Soubiran}, {Tanga}, {Walton},
  {Bailer-Jones}, {Bastian}, {Drimmel}, {Jansen}, {Katz}, {van Leeuwen},
  {Bakker}, {Cacciari}, {Casta{\~n}eda}, {De Angeli}, {Fabricius}, {Fouesneau},
  {Fr{\'e}mat}, {Galluccio}, {Guerrier}, {Heiter}, {Masana}, {Messineo},
  {Nicolas}, {Nienartowicz}, {Pailler}, {Riclet}, {Roux}, {Seabroke}, {Sordo},
  {Th{\'e}venin}, {Gracia-Abril}, {Portell}, {Altmann}, {Andrae}, {Audard},
  {Bellas-Velidis}, {Benson}, {Berthier}, {Blomme}, {Burgess}, {Busonero},
  {Busso}, {C{\'a}novas}, {Carry}, {Cellino}, {Cheek}, {Clementini},
  {Davidson}, {de Teodoro}, {Nu{\~n}ez Campos}, {Delchambre}, {Dell'Oro},
  {Esquej}, {Fern{\'a}ndez-Hern{\'a}ndez}, {Fraile}, {Garabato},
  {Garc{\'\i}a-Lario}, {Haigron}, {Hambly}, {Harrison}, {Hern{\'a}ndez},
  {Hestroffer}, {Hodgkin}, {Jan{\ss}en}, {Jevardat de Fombelle}, {Jordan},
  {Krone-Martins}, {Lanzafame}, {L{\"o}ffler}, {Marchal}, {Marrese},
  {Moitinho}, {Muinonen}, {Osborne}, {Pancino}, {Pauwels}, {Recio-Blanco},
  {Reyl{\'e}}, {Riello}, {Rimoldini}, {Roegiers}, {Rybizki}, {Sarro}, {Smith},
  {Utrilla}, {van Leeuwen}, {Abbas}, {{\'A}brah{\'a}m}, {Abreu Aramburu},
  {Aerts}, {Aguado}, {Ajaj}, {Aldea-Montero}, {Altavilla}, {{\'A}lvarez},
  {Alves}, {Anders}, {Anderson}, {Anglada Varela}, {Antoja}, {Baines}, {Baker},
  {Balaguer-N{\'u}{\~n}ez}, {Balbinot}, {Balog}, {Barache}, {Barbato},
  {Barros}, {Bartolom{\'e}}, {Bassilana}, {Becciani}, {Bellazzini},
  {Berihuete}, {Bernet}, {Bertone}, {Bianchi}, {Binnenfeld}, {Blanco-Cuaresma},
  {Blazere}, {Boch}, {Bombrun}, {Bossini}, {Bouquillon}, {Bragaglia},
  {Bramante}, {Breedt}, {Bressan}, {Brouillet}, {Brugaletta}, {Bucciarelli},
  {Burlacu}, {Butkevich}, {Buzzi}, {Caffau}, {Cancelliere}, {Cantat-Gaudin},
  {Carballo}, {Carlucci}, {Carnerero}, {Carrasco}, {Casamiquela}, {Castellani},
  {Castro-Ginard}, {Chaoul}, {Charlot}, {Chemin}, {Chiaramida}, {Chiavassa},
  {Chornay}, {Comoretto}, {Contursi}, {Cooper}, {Cornez}, {Cowell}, {Crifo},
  {Cropper}, {Crosta}, {Crowley}, {Dafonte}, {Dapergolas}, {David}, {de
  Laverny}, {De Luise}, {De March}, {De Ridder}, {de Souza}, {de Torres}, {del
  Peloso}, {del Pozo}, {Delbo}, {Delgado}, {Delisle}, {Demouchy},
  {Dharmawardena}, {Diakite}, {Diener}, {Distefano}, {Dolding}, {Enke},
  {Fabre}, {Fabrizio}, {Fedorets}, {Fernique}, {Figueras}, {Fournier},
  {Fouron}, {Fragkoudi}, {Gai}, {Garcia-Gutierrez}, {Garcia-Reinaldos},
  {Garc{\'\i}a-Torres}, {Garofalo}, {Gavel}, {Gerlach}, {Geyer}, {Gilmore},
  {Girona}, {Giuffrida}, {Gomel}, {Gomez}, {Gonz{\'a}lez-N{\'u}{\~n}ez},
  {Gonz{\'a}lez-Santamar{\'\i}a}, {Gonz{\'a}lez-Vidal}, {Granvik}, {Guillout},
  {Guiraud}, {Guti{\'e}rrez-S{\'a}nchez}, {Guy}, {Hatzidimitriou}, {Hauser},
  {Haywood}, {Helmer}, {Helmi}, {Sarmiento}, {Hidalgo}, {Hilger},
  {H{\l}adczuk}, {Hobbs}, {Holland}, {Huckle}, {Jardine}, {Jasniewicz},
  {Jean-Antoine Piccolo}, {Jim{\'e}nez-Arranz}, {Juaristi Campillo}, {Julbe},
  {Karbevska}, {Khanna}, {Kordopatis}, {Korn}, {K{\'o}sp{\'a}l},
  {Kostrzewa-Rutkowska}, {Kruszy{\'n}ska}, {Kun}, {Laizeau}, {Lambert},
  {Lanza}, {Lasne}, {Le Campion}, {Lebreton}, {Lebzelter}, {Leccia},
  {Lecoeur-Taibi}, {Liao}, {Licata}, {Lindstr{\o}m}, {Lister}, {Livanou},
  {Lobel}, {Lorca}, {Loup}, {Madrero Pardo}, {Magdaleno Romeo}, {Managau},
  {Mann}, {Manteiga}, {Marchant}, {Marconi}, {Marcos}, {Marcos Santos},
  {Mar{\'\i}n Pina}, {Marinoni}, {Marocco}, {Marshall}, {Martin Polo},
  {Mart{\'\i}n-Fleitas}, {Marton}, {Mary}, {Masip}, {Massari},
  {Mastrobuono-Battisti}, {McMillan}, {Messina}, {Michalik}, {Millar}, {Mints},
  {Molina}, {Molinaro}, {Moln{\'a}r}, {Monari}, {Mongui{\'o}}, {Montegriffo},
  {Montero}, {Mor}, {Mora}, {Morbidelli}, {Morris}, {Muraveva}, {Murphy},
  {Musella}, {Nagy}, {Noval}, {Oca{\~n}a}, {Ogden}, {Ordenovic}, {Osinde},
  {Pagani}, {Pagano}, {Palaversa}, {Palicio}, {Pallas-Quintela}, {Panahi},
  {Payne-Wardenaar}, {Pe{\~n}alosa Esteller}, {Penttil{\"a}}, {Pichon},
  {Piersimoni}, {Pineau}, {Plachy}, {Plum}, {Poggio}, {Pr{\v{s}}a}, {Pulone},
  {Racero}, {Ragaini}, {Rainer}, {Raiteri}, {Ramos}, {Ramos-Lerate}, {Regibo},
  {Richards}, {Rios Diaz}, {Ripepi}, {Riva}, {Rix}, {Rixon}, {Robichon},
  {Robin}, {Robin}, {Roelens}, {Rogues}, {Rohrbasser}, {Romero-G{\'o}mez},
  {Rowell}, {Royer}, {Ruz Mieres}, {Rybicki}, {S{\'a}ez N{\'u}{\~n}ez},
  {Sagrist{\`a} Sell{\'e}s}, {Salguero}, {Samaras}, {Sanchez Gimenez}, {Sanna},
  {Santove{\~n}a}, {Sarasso}, {Schultheis}, {Sciacca}, {Segol}, {Segovia},
  {Semeux}, {Siddiqui}, {Siebert}, {Siltala}, {Silvelo}, {Slezak}, {Slezak},
  {Smart}, {Snaith}, {Solano}, {Solitro}, {Souami}, {Souchay}, {Spagna},
  {Spina}, {Spoto}, {Steele}, {Steidelm{\"u}ller}, {Stephenson}, {S{\"u}veges},
  {Surdej}, {Szabados}, {Szegedi-Elek}, {Taris}, {Taylor}, {Teixeira},
  {Tolomei}, {Tonello}, {Torra}, {Torra}, {Torralba Elipe}, {Trabucchi},
  {Tsounis}, {Turon}, {Ulla}, {Unger}, {Vaillant}, {van Dillen}, {van Reeven},
  {Vanel}, {Vecchiato}, {Viala}, {Vicente}, {Voutsinas}, {Weiler}, {Wevers},
  {Wyrzykowski}, {Yoldas}, {Yvard}, {Zhao}, {Zorec}, \& {Zucker}}]{Gaia}
{Gaia Collaboration}, {Arenou}, F., {Babusiaux}, C., {et~al.} 2023, \aap, 674,
  A34, \dodoi{10.1051/0004-6361/202243782}

\bibitem[{{Georgakarakos}(2013)}]{2013Georga}
{Georgakarakos}, N. 2013, \na, 23, 41, \dodoi{10.1016/j.newast.2013.02.004}

\bibitem[{{Giacobbo} \& {Mapelli}(2020)}]{Giacobbo2020}
{Giacobbo}, N., \& {Mapelli}, M. 2020, \apj, 891, 141,
  \dodoi{10.3847/1538-4357/ab7335}

\bibitem[{Gond{\'a}n {et~al.}(2018)Gond{\'a}n, Kocsis, Raffai, \&
  Frei}]{gondan2018eccentric}
Gond{\'a}n, L., Kocsis, B., Raffai, P., \& Frei, Z. 2018, ApJ, 860, 5

\bibitem[{{Grishin} \& {Perets}(2022)}]{grishin2022}
{Grishin}, E., \& {Perets}, H.~B. 2022, \mnras, 512, 4993,
  \dodoi{10.1093/mnras/stac706}

\bibitem[{{Grishin} {et~al.}(2018){Grishin}, {Perets}, \&
  {Fragione}}]{2018grishin}
{Grishin}, E., {Perets}, H.~B., \& {Fragione}, G. 2018, \mnras, 481, 4907,
  \dodoi{10.1093/mnras/sty2477}

\bibitem[{{Grishin} {et~al.}(2017){Grishin}, {Perets}, {Zenati}, \&
  {Michaely}}]{grishin_stab}
{Grishin}, E., {Perets}, H.~B., {Zenati}, Y., \& {Michaely}, E. 2017, \mnras,
  466, 276, \dodoi{10.1093/mnras/stw3096}

\bibitem[{{Ha} {et~al.}(2022){Ha}, {Li}, {Kounkel}, {Xu}, {Li}, \&
  {Zheng}}]{Ha22}
{Ha}, T., {Li}, Y., {Kounkel}, M., {et~al.} 2022, \apj, 934, 7,
  \dodoi{10.3847/1538-4357/ac76bf}

\bibitem[{{Ha} {et~al.}(2021){Ha}, {Li}, {Xu}, {Kounkel}, \& {Li}}]{Ha21}
{Ha}, T., {Li}, Y., {Xu}, S., {Kounkel}, M., \& {Li}, H. 2021, \apjl, 907, L40,
  \dodoi{10.3847/2041-8213/abd8c9}

\bibitem[{Hamers(2020)}]{Hamers}
Hamers, A.~S. 2020, MNRAS, 494, 5492, \dodoi{10.1093/mnras/staa1084}

\bibitem[{{Hamers} {et~al.}(2022){Hamers}, {Perets}, {Thompson}, \&
  {Neunteufel}}]{hamers2022}
{Hamers}, A.~S., {Perets}, H.~B., {Thompson}, T.~A., \& {Neunteufel}, P. 2022,
  \apj, 925, 178, \dodoi{10.3847/1538-4357/ac400b}

\bibitem[{{Harada} {et~al.}(2021){Harada}, {Hirano}, {Machida}, \&
  {Hosokawa}}]{Hara21}
{Harada}, N., {Hirano}, S., {Machida}, M.~N., \& {Hosokawa}, T. 2021, \mnras,
  508, 3730, \dodoi{10.1093/mnras/stab2780}

\bibitem[{{Hayashi} {et~al.}(2022){Hayashi}, {Trani}, \&
  {Suto}}]{hayashi_stab1}
{Hayashi}, T., {Trani}, A.~A., \& {Suto}, Y. 2022, \apj, 939, 81,
  \dodoi{10.3847/1538-4357/ac8f48}

\bibitem[{{Hayashi} {et~al.}(2023){Hayashi}, {Trani}, \&
  {Suto}}]{hayashi_stab2}
---. 2023, \apj, 943, 58, \dodoi{10.3847/1538-4357/acac1e}

\bibitem[{{Hennebelle} \& {Ciardi}(2009)}]{Henn09}
{Hennebelle}, P., \& {Ciardi}, A. 2009, \aap, 506, L29,
  \dodoi{10.1051/0004-6361/200913008}

\bibitem[{Hills(1983)}]{hills1983effects}
Hills, J. 1983, Astrophysical Journal, Part 1, vol. 267, Apr. 1, 1983, p.
  322-333. Research supported by the US Department of Energy., 267, 322

\bibitem[{Hoang {et~al.}(2018)Hoang, Naoz, Kocsis, Rasio, \&
  Dosopoulou}]{hoang2018black}
Hoang, B.-M., Naoz, S., Kocsis, B., Rasio, F.~A., \& Dosopoulou, F. 2018, ApJ,
  856, 140

\bibitem[{{Hu} {et~al.}(2019){Hu}, {Yuen}, {Lazarian}, {Ho}, {Benjamin},
  {Hill}, {Lockman}, {Goldsmith}, \& {Lazarian}}]{Humc19}
{Hu}, Y., {Yuen}, K.~H., {Lazarian}, V., {et~al.} 2019, Nature Astronomy, 3,
  776, \dodoi{10.1038/s41550-019-0769-0}

\bibitem[{{Hwang}(2023)}]{Hwa23}
{Hwang}, H.-C. 2023, \mnras, 518, 1750, \dodoi{10.1093/mnras/stac3116}

\bibitem[{Hwang {et~al.}(2022)Hwang, Ting, \& Zakamska}]{hwang2022}
Hwang, H.-C., Ting, Y.-S., \& Zakamska, N.~L. 2022, MNRAS, 512, 3383

\bibitem[{{Joos} {et~al.}(2012){Joos}, {Hennebelle}, \& {Ciardi}}]{Joos12}
{Joos}, M., {Hennebelle}, P., \& {Ciardi}, A. 2012, \aap, 543, A128,
  \dodoi{10.1051/0004-6361/201118730}

\bibitem[{{Kimball} {et~al.}(2023){Kimball}, {Imperato}, {Kalogera}, {Rocha},
  {Doctor}, {Andrews}, {Dotter}, {Zapartas}, {Bavera}, {Kovlakas}, {Fragos},
  {Srivastava}, {Misra}, {Sun}, \& {Xing}}]{kimball2023kick}
{Kimball}, C., {Imperato}, S., {Kalogera}, V., {et~al.} 2023, \apjl, 952, L34,
  \dodoi{10.3847/2041-8213/ace526}

\bibitem[{{Kiseleva} {et~al.}(1996){Kiseleva}, {Aarseth}, {Eggleton}, \& {de La
  Fuente Marcos}}]{kiseleva}
{Kiseleva}, L.~G., {Aarseth}, S.~J., {Eggleton}, P.~P., \& {de La Fuente
  Marcos}, R. 1996, in Astronomical Society of the Pacific Conference Series,
  Vol.~90, The Origins, Evolution, and Destinies of Binary Stars in Clusters,
  ed. E.~F. {Milone} \& J.~C. {Mermilliod}, 433

\bibitem[{Kozai(1962)}]{kozai}
Kozai, Y. 1962, AJ, 67, 591

\bibitem[{{Krolikowski} {et~al.}(2021){Krolikowski}, {Kraus}, \&
  {Rizzuto}}]{Krol21}
{Krolikowski}, D.~M., {Kraus}, A.~L., \& {Rizzuto}, A.~C. 2021, \aj, 162, 110,
  \dodoi{10.3847/1538-3881/ac0632}

\bibitem[{{Kummer} {et~al.}(2023){Kummer}, {Toonen}, \& {de
  Koter}}]{kummer2023}
{Kummer}, F., {Toonen}, S., \& {de Koter}, A. 2023, \aap, 678, A60,
  \dodoi{10.1051/0004-6361/202347179}

\bibitem[{{Lalande} \& {Trani}(2022)}]{lalande_stab}
{Lalande}, F., \& {Trani}, A.~A. 2022, \apj, 938, 18,
  \dodoi{10.3847/1538-4357/ac8eab}

\bibitem[{{Lei} {et~al.}(2018){Lei}, {Circi}, \& {Ortore}}]{lei2018}
{Lei}, H., {Circi}, C., \& {Ortore}, E. 2018, \mnras, 481, 4602,
  \dodoi{10.1093/mnras/sty2619}

\bibitem[{Leigh {et~al.}(2018)Leigh, Geller, McKernan, Ford, Mac~Low,
  Bellovary, Haiman, Lyra, Samsing, O'Dowd, {et~al.}}]{leigh2018rate}
Leigh, N.~W., Geller, A.~M., McKernan, B., {et~al.} 2018, MNRAS, 474, 5672

\bibitem[{{Li} {et~al.}(2022){Li}, {Lai}, \& {Rodet}}]{li2022b}
{Li}, J., {Lai}, D., \& {Rodet}, L. 2022, \apj, 934, 154,
  \dodoi{10.3847/1538-4357/ac7c0d}

\bibitem[{{Li} \& {Lai}(2022)}]{li2022}
{Li}, R., \& {Lai}, D. 2022, \mnras, 517, 1602, \dodoi{10.1093/mnras/stac2577}

\bibitem[{{Li} {et~al.}(2021){Li}, {Dempsey}, {Li}, {Li}, \& {Li}}]{li2021}
{Li}, Y.-P., {Dempsey}, A.~M., {Li}, S., {Li}, H., \& {Li}, J. 2021, \apj, 911,
  124, \dodoi{10.3847/1538-4357/abed48}

\bibitem[{{Li} {et~al.}(2013){Li}, {Krasnopolsky}, \& {Shang}}]{Liz13}
{Li}, Z.-Y., {Krasnopolsky}, R., \& {Shang}, H. 2013, \apj, 774, 82,
  \dodoi{10.1088/0004-637X/774/1/82}

\bibitem[{Lidov(1962)}]{lidov}
Lidov, M.~L. 1962, Planetary and Space Science, 9, 719

\bibitem[{Lipunov {et~al.}(1997)Lipunov, Postnov, \&
  Prokhorov}]{lipunov1997black}
Lipunov, V., Postnov, K., \& Prokhorov, M. 1997, Astronomy Letters, 23, 492

\bibitem[{Lipunov {et~al.}(2017)Lipunov, Kornilov, Gorbovskoy, Buckley,
  Tiurina, Balanutsa, Kuznetsov, Greiner, Vladimirov, Vlasenko,
  {et~al.}}]{lipunov2017first}
Lipunov, V., Kornilov, V., Gorbovskoy, E., {et~al.} 2017, MNRAS, 465, 3656

\bibitem[{Lithwick \& Naoz(2011)}]{lithwick2011eccentric}
Lithwick, Y., \& Naoz, S. 2011, ApJ, 742, 94

\bibitem[{Liu \& Lai(2017)}]{LL17}
Liu, B., \& Lai, D. 2017, ApJL, 846, L11

\bibitem[{Liu \& Lai(2018)}]{LL18}
---. 2018, ApJ, 863, 68

\bibitem[{{Liu} \& {Lai}(2019)}]{bin_misc5}
{Liu}, B., \& {Lai}, D. 2019, MNRAS, 483, 4060, \dodoi{10.1093/mnras/sty3432}

\bibitem[{{Liu} \& {Lai}(2020)}]{bin_misc2}
---. 2020, Phys. Rev. D, 102, 023020, \dodoi{10.1103/PhysRevD.102.023020}

\bibitem[{Liu \& Lai(2021)}]{liu2021hierarchical}
Liu, B., \& Lai, D. 2021, MNRAS, 502, 2049

\bibitem[{{Liu} {et~al.}(2019{\natexlab{a}}){Liu}, {Lai}, \& {Wang}}]{LL19}
{Liu}, B., {Lai}, D., \& {Wang}, Y.-H. 2019{\natexlab{a}}, ApJ, 881, 41,
  \dodoi{10.3847/1538-4357/ab2dfb}

\bibitem[{{Liu} {et~al.}(2019{\natexlab{b}}){Liu}, {Lai}, \& {Wang}}]{LLW_apjl}
---. 2019{\natexlab{b}}, ApJL, 883, L7, \dodoi{10.3847/2041-8213/ab40c0}

\bibitem[{Liu {et~al.}(2015)Liu, Mu{\~n}oz, \& Lai}]{LML15}
Liu, B., Mu{\~n}oz, D.~J., \& Lai, D. 2015, MNRAS, 447, 747

\bibitem[{{Lund} \& {Bonnell}(2018)}]{Lund18}
{Lund}, K., \& {Bonnell}, I.~A. 2018, \mnras, 479, 2235,
  \dodoi{10.1093/mnras/sty1584}

\bibitem[{{Luo} {et~al.}(2016){Luo}, {Katz}, \& {Dong}}]{luo2016}
{Luo}, L., {Katz}, B., \& {Dong}, S. 2016, \mnras, 458, 3060,
  \dodoi{10.1093/mnras/stw475}

\bibitem[{Mandel \& De~Mink(2016)}]{mandel2016merging}
Mandel, I., \& De~Mink, S.~E. 2016, MNRAS, 458, 2634

\bibitem[{{Mandel} \& {M{\"u}ller}(2020)}]{mandel2020kick}
{Mandel}, I., \& {M{\"u}ller}, B. 2020, \mnras, 499, 3214,
  \dodoi{10.1093/mnras/staa3043}

\bibitem[{{Mangipudi} {et~al.}(2022){Mangipudi}, {Grishin}, {Trani}, \&
  {Mandel}}]{mangipudi2022}
{Mangipudi}, A., {Grishin}, E., {Trani}, A.~A., \& {Mandel}, I. 2022, \apj,
  934, 44, \dodoi{10.3847/1538-4357/ac7958}

\bibitem[{Marchant {et~al.}(2016)Marchant, Langer, Podsiadlowski, Tauris, \&
  Moriya}]{marchant2016new}
Marchant, P., Langer, N., Podsiadlowski, P., Tauris, T.~M., \& Moriya, T.~J.
  2016, A\&A, 588, A50

\bibitem[{Mardling \& Aarseth(2001)}]{mardling2001tidal}
Mardling, R.~A., \& Aarseth, S.~J. 2001, MNRAS, 321, 398

\bibitem[{{Martinez} {et~al.}(2022){Martinez}, {Rodriguez}, \&
  {Fragione}}]{martinez2021mass}
{Martinez}, M. A.~S., {Rodriguez}, C.~L., \& {Fragione}, G. 2022, \apj, 937,
  78, \dodoi{10.3847/1538-4357/ac8d55}

\bibitem[{McKernan {et~al.}(2012)McKernan, Ford, Lyra, \&
  Perets}]{mckernan2012}
McKernan, B., Ford, K. E.~S., Lyra, W., \& Perets, H.~B. 2012, MNRAS, 425, 460,
  \dodoi{10.1111/j.1365-2966.2012.21486.x}

\bibitem[{Miller \& Hamilton(2002)}]{miller2002four}
Miller, M.~C., \& Hamilton, D.~P. 2002, ApJ, 576, 894

\bibitem[{Miller \& Lauburg(2009)}]{miller2009mergers}
Miller, M.~C., \& Lauburg, V.~M. 2009, ApJ, 692, 917

\bibitem[{Naoz(2016)}]{naoz2016eccentric}
Naoz, S. 2016, ARAA, 54, 441

\bibitem[{O'leary {et~al.}(2006)O'leary, Rasio, Fregeau, Ivanova, \&
  O'Shaughnessy}]{o2006binary}
O'leary, R.~M., Rasio, F.~A., Fregeau, J.~M., Ivanova, N., \& O'Shaughnessy, R.
  2006, ApJ, 637, 937

\bibitem[{Podsiadlowski {et~al.}(2003)Podsiadlowski, Rappaport, \&
  Han}]{podsiadlowski2003formation}
Podsiadlowski, P., Rappaport, S., \& Han, Z. 2003, MNRAS, 341, 385

\bibitem[{{Portegies Zwart} \& {McMillan}(2000)}]{zwart1999black}
{Portegies Zwart}, S.~F., \& {McMillan}, S. L.~W. 2000, \apjl, 528, L17,
  \dodoi{10.1086/312422}

\bibitem[{Randall \& Xianyu(2018{\natexlab{a}})}]{randall2018induced}
Randall, L., \& Xianyu, Z.-Z. 2018{\natexlab{a}}, ApJ, 853, 93

\bibitem[{Randall \& Xianyu(2018{\natexlab{b}})}]{randall2018analytical}
---. 2018{\natexlab{b}}, ApJ, 864, 134

\bibitem[{{Repetto} {et~al.}(2012){Repetto}, {Davies}, \&
  {Sigurdsson}}]{repetto2012}
{Repetto}, S., {Davies}, M.~B., \& {Sigurdsson}, S. 2012, \mnras, 425, 2799,
  \dodoi{10.1111/j.1365-2966.2012.21549.x}

\bibitem[{Riley {et~al.}(2021)Riley, Mandel, Marchant, Butler, Nathaniel,
  Neijssel, Shortt, \& Vigna-G{\'o}mez}]{riley2021chemically}
Riley, J., Mandel, I., Marchant, P., {et~al.} 2021, MNRAS, 505, 663

\bibitem[{Rodriguez {et~al.}(2018)Rodriguez, Amaro-Seoane, Chatterjee, \&
  Rasio}]{rodriguez2018post}
Rodriguez, C.~L., Amaro-Seoane, P., Chatterjee, S., \& Rasio, F.~A. 2018, Phys.
  Rev. Lett., 120, 151101

\bibitem[{Rodriguez {et~al.}(2015)Rodriguez, Morscher, Pattabiraman,
  Chatterjee, Haster, \& Rasio}]{rodriguez2015binary}
Rodriguez, C.~L., Morscher, M., Pattabiraman, B., {et~al.} 2015, Phys. Rev.
  Lett., 115, 051101

\bibitem[{Samsing \& D’Orazio(2018)}]{samsing2018black}
Samsing, J., \& D’Orazio, D.~J. 2018, MNRAS, 481, 5445

\bibitem[{Samsing \& Ramirez-Ruiz(2017)}]{samsing2017assembly}
Samsing, J., \& Ramirez-Ruiz, E. 2017, ApJL, 840, L14

\bibitem[{Samsing {et~al.}(2022)Samsing, Bartos, D’Orazio, Haiman, Kocsis,
  Leigh, Liu, Pessah, \& Tagawa}]{samsing2022agn}
Samsing, J., Bartos, I., D’Orazio, D., {et~al.} 2022, Nature, 603, 237

\bibitem[{{Secunda} {et~al.}(2019){Secunda}, {Bellovary}, {Mac Low}, {Ford},
  {McKernan}, {Leigh}, {Lyra}, \& {S{\'a}ndor}}]{secunda2019}
{Secunda}, A., {Bellovary}, J., {Mac Low}, M.-M., {et~al.} 2019, \apj, 878, 85,
  \dodoi{10.3847/1538-4357/ab20ca}

\bibitem[{Silsbee \& Tremaine(2017)}]{silsbee2017lidov}
Silsbee, K., \& Tremaine, S. 2017, ApJ, 836, 39

\bibitem[{Spera \& Mapelli(2017)}]{spera2017very}
Spera, M., \& Mapelli, M. 2017, MNRAS, 470, 4739

\bibitem[{Spera {et~al.}(2015)Spera, Mapelli, \& Bressan}]{spera2015mass}
Spera, M., Mapelli, M., \& Bressan, A. 2015, Monthly Notices of the Royal
  Astronomical Society, 451, 4086

\bibitem[{{Stone} {et~al.}(2017){Stone}, {Metzger}, \& {Haiman}}]{stone2017}
{Stone}, N.~C., {Metzger}, B.~D., \& {Haiman}, Z. 2017, \mnras, 464, 946,
  \dodoi{10.1093/mnras/stw2260}

\bibitem[{Su {et~al.}(2021{\natexlab{a}})Su, Lai, \& Liu}]{su2020spin}
Su, Y., Lai, D., \& Liu, B. 2021{\natexlab{a}}, Phys. Rev. D, 103, 063040

\bibitem[{Su {et~al.}(2021{\natexlab{b}})Su, Liu, \& Lai}]{su2021mass}
Su, Y., Liu, B., \& Lai, D. 2021{\natexlab{b}}, MNRAS, 505, 3681

\bibitem[{Tagawa {et~al.}(2020)Tagawa, Haiman, \& Kocsis}]{tagawa2020formation}
Tagawa, H., Haiman, Z., \& Kocsis, B. 2020, ApJ, 898, 25

\bibitem[{{Tokovinin}(2020)}]{Toko20}
{Tokovinin}, A. 2020, \mnras, 496, 987, \dodoi{10.1093/mnras/staa1639}

\bibitem[{{Toonen} {et~al.}(2022){Toonen}, {Boekholt}, \& {Portegies
  Zwart}}]{toonen2022}
{Toonen}, S., {Boekholt}, T.~C.~N., \& {Portegies Zwart}, S. 2022, \aap, 661,
  A61, \dodoi{10.1051/0004-6361/202141991}

\bibitem[{{Tory} {et~al.}(2022){Tory}, {Grishin}, \& {Mandel}}]{2022tory}
{Tory}, M., {Grishin}, E., \& {Mandel}, I. 2022, \pasa, 39, e062,
  \dodoi{10.1017/pasa.2022.57}

\bibitem[{{Trani} {et~al.}(2022){Trani}, {Rastello}, {Di Carlo},
  {Santoliquido}, {Tanikawa}, \& {Mapelli}}]{trani2022}
{Trani}, A.~A., {Rastello}, S., {Di Carlo}, U.~N., {et~al.} 2022, \mnras, 511,
  1362, \dodoi{10.1093/mnras/stac122}

\bibitem[{{Tremaine}(2023)}]{tremaine2023brown}
{Tremaine}, S. 2023, \mnras, 522, 937, \dodoi{10.1093/mnras/stad1029}

\bibitem[{{Tsukamoto} {et~al.}(2018){Tsukamoto}, {Okuzumi}, {Iwasaki},
  {Machida}, \& {Inutsuka}}]{Tsuka18}
{Tsukamoto}, Y., {Okuzumi}, S., {Iwasaki}, K., {Machida}, M.~N., \& {Inutsuka},
  S. 2018, \apj, 868, 22, \dodoi{10.3847/1538-4357/aae4dc}

\bibitem[{{Vigna-G{\'o}mez} {et~al.}(2024){Vigna-G{\'o}mez}, {Willcox},
  {Tamborra}, {Mandel}, {Renzo}, {Wagg}, {Janka}, {Kresse}, {Bodensteiner},
  {Shenar}, \& {Tauris}}]{vignagomez2023natal}
{Vigna-G{\'o}mez}, A., {Willcox}, R., {Tamborra}, I., {et~al.} 2024, \prl, 132,
  191403, \dodoi{10.1103/PhysRevLett.132.191403}

\bibitem[{von Zeipel(1910)}]{zeipel}
von Zeipel, H. 1910, Astronomische Nachrichten, 183, 345

\bibitem[{{Vynatheya} {et~al.}(2022){Vynatheya}, {Hamers}, {Mardling}, \&
  {Bellinger}}]{vynatheya_stab}
{Vynatheya}, P., {Hamers}, A.~S., {Mardling}, R.~A., \& {Bellinger}, E.~P.
  2022, \mnras, 516, 4146, \dodoi{10.1093/mnras/stac2540}

\bibitem[{Wen(2003)}]{wen2003eccentricity}
Wen, L. 2003, ApJ, 598, 419

\bibitem[{Xu {et~al.}(2023)Xu, Hwang, Hamilton, \& Lai}]{xu2023wide}
Xu, S., Hwang, H.-C., Hamilton, C., \& Lai, D. 2023, ApJL, 949, L28

\bibitem[{{Xu} \& {Lazarian}(2020)}]{Xudyn20}
{Xu}, S., \& {Lazarian}, A. 2020, \apj, 899, 115,
  \dodoi{10.3847/1538-4357/aba7ba}

\bibitem[{{Yuen} {et~al.}(2022){Yuen}, {Ho}, {Law}, {Chen}, \&
  {Lazarian}}]{Yuen22}
{Yuen}, K.~H., {Ho}, K.~W., {Law}, C.~Y., {Chen}, A., \& {Lazarian}, A. 2022,
  arXiv e-prints, arXiv:2204.13760, \dodoi{10.48550/arXiv.2204.13760}

\bibitem[{{Zhou} {et~al.}(2022){Zhou}, {Li}, \& {Chen}}]{Zhou21}
{Zhou}, J.-X., {Li}, G.-X., \& {Chen}, B.-Q. 2022, \mnras, 513, 638,
  \dodoi{10.1093/mnras/stac900}

\bibitem[{Ziosi {et~al.}(2014)Ziosi, Mapelli, Branchesi, \&
  Tormen}]{ziosi2014dynamics}
Ziosi, B.~M., Mapelli, M., Branchesi, M., \& Tormen, G. 2014, MNRAS, 441, 3703

\end{thebibliography}
\bibliographystyle{aasjournal}




\appendix
\section{Kick in the triples: dependence on the initial \texorpdfstring{$e_\OUT^0$}{eoutk}}\label{Appendix A}

\begin{figure}
\centering
\begin{tabular}{cccc}
\includegraphics[width=8cm]{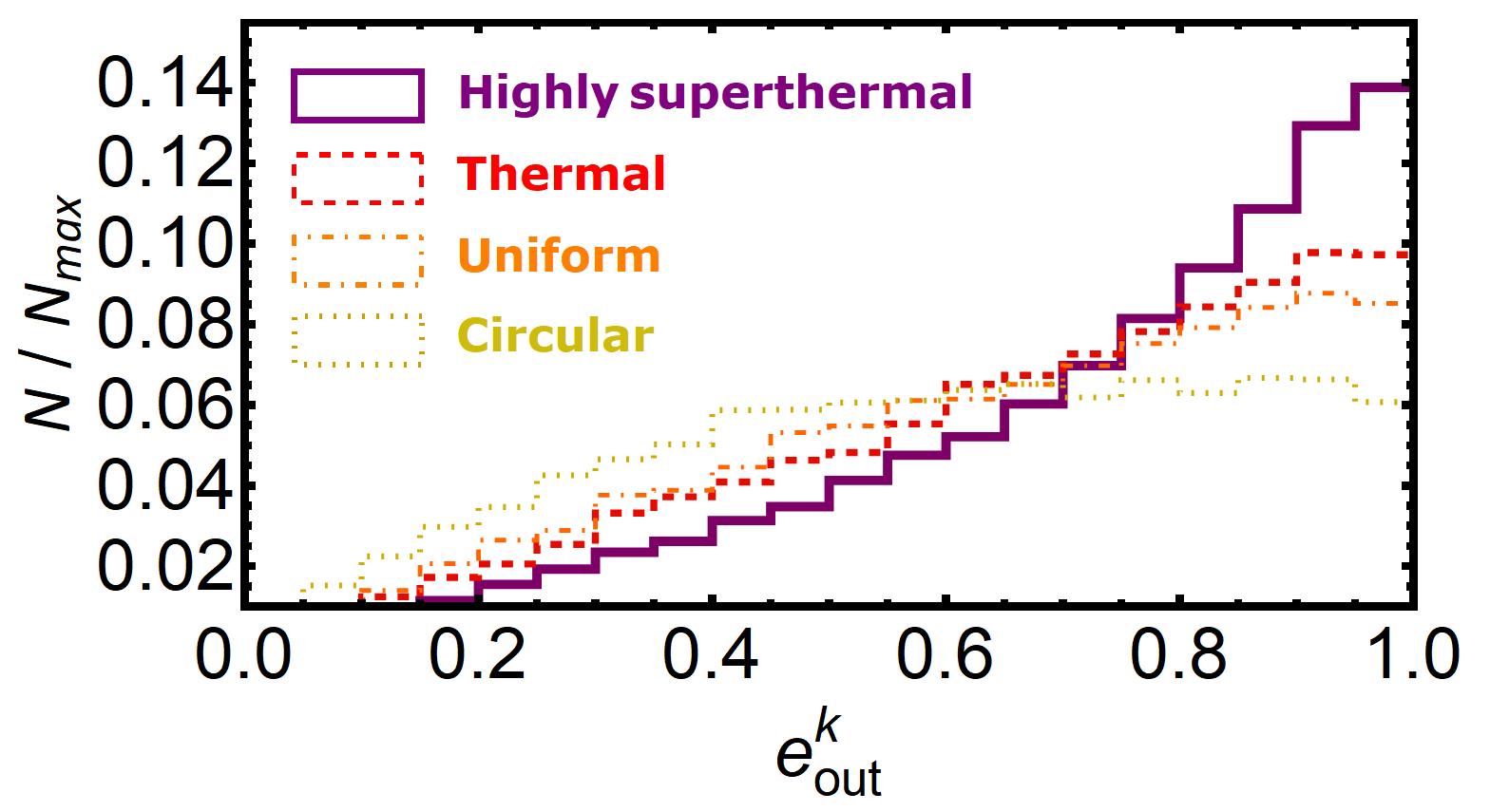}
\end{tabular}
\caption{The probability distributions of post-SN eccentricities of extremely hierarchical triples, in which
we consider $a_\IN^0=10\;\mathrm{AU}$ and $a_\OUT^0=10^4\;\mathrm{AU}$. We include four different $e_\OUT^0$ distributions in our calculations,
including $e_\OUT^0=0$ (circular), uniform distribution over $[0, 1]$ (uniform), $f(e_\OUT^0)\propto(e_\OUT^0)$ (thermal)
and distributions given by Eq. (\ref{eq:pe}) (highly superthermal).
}\label{fig:A1}
\end{figure}

Here, we examine how the initial eccentricities affect the post-kick eccentricity distributions.
Similar to the top panel of Fig.~\ref{fig:Survival fraction}, we focus on the sufficiently hierarchical triples
($a_\IN^0/a_\OUT^0=0.001$) and consider a variety of initial $e_\OUT^0$.
As shown in Fig.~\ref{fig:A1}, a full range of $e_\OUT^k$
can be always produced regardless of the initial distributions, indicating that the memory of the initial geometry is likely to be lost for such
extreme hierarchical triple system.
As discussed in the main text, when the system becomes less hierarchical, the effect of the kick can be reduced.

\begin{figure}
\centering
\begin{tabular}{cccc}
\includegraphics[width=8cm]{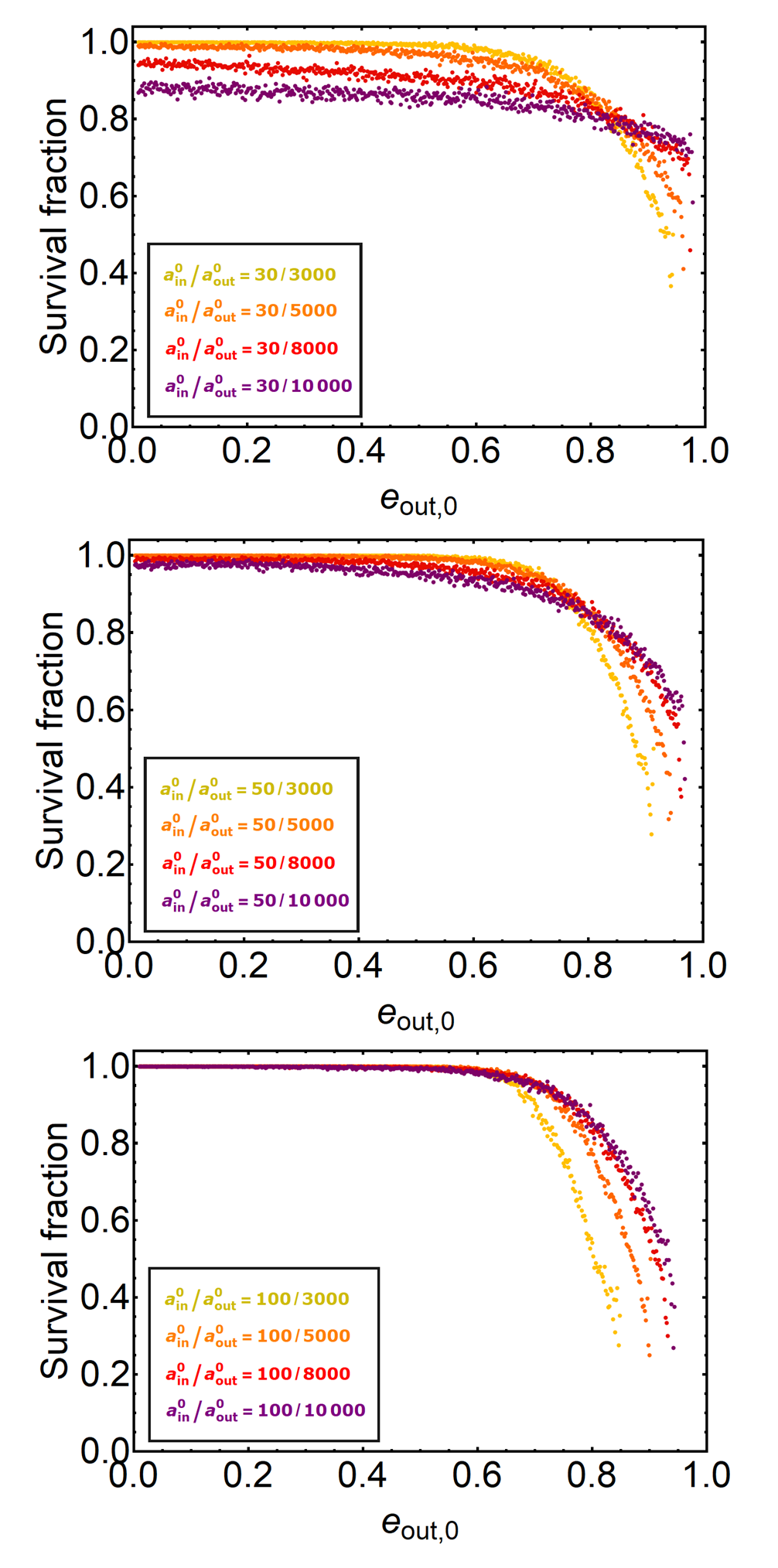}
\end{tabular}
\caption{Survival fraction as a function of the initial eccentricity of the outer binary.
System parameters are $m_1^0=m_3^0=33.3M_\odot$, $m_2^0=22.2M_\odot$, $e_\IN^0=0.001$,
$a_\IN^0=30\;\mathrm{AU}$ (top panel), $a_\IN^0=50\;\mathrm{AU}$ (middle panel)
and $a_\IN^0=1000\;\mathrm{AU}$ (bottom panel).
Different initial semimajor axes of the outer binary are considered for each $a_\IN^0$ (as labeled).
Here, for each initial $e_\OUT^0$, the system is evolved until it undergoes three SNs.
To cover all possible geometry of the triples, we run $10^3$ simulations and calculate the mean value of the survival fractions.
}\label{fig:A2}
\end{figure}

To explore the dependence of the survival fraction on the initial eccentricity distribution,
we consider a wide range of $a_\IN^0/a_\OUT^0$ and perform the numerical integrations with dense grids of $e_\OUT^0$.
The results are shown in Fig.~\ref{fig:A2}.
We see that in general,  $F_\mathrm{survival}$ appears to be constant when $e_\OUT^0\lesssim0.6$ and
decreases dramatically when $e_\OUT^0\gtrsim0.6$.
As the level of hierarchy $a_\IN^0/a_\OUT^0$ increases,
the survival fraction tends to decrease with a lower $e_\OUT^0$.
This is mainly due to the stability.

\section{Effect of Mass Loss on Outer Eccentricity Distribution}\label{app:eout}

\begin{figure}
    \centering
    \includegraphics[width=\columnwidth]{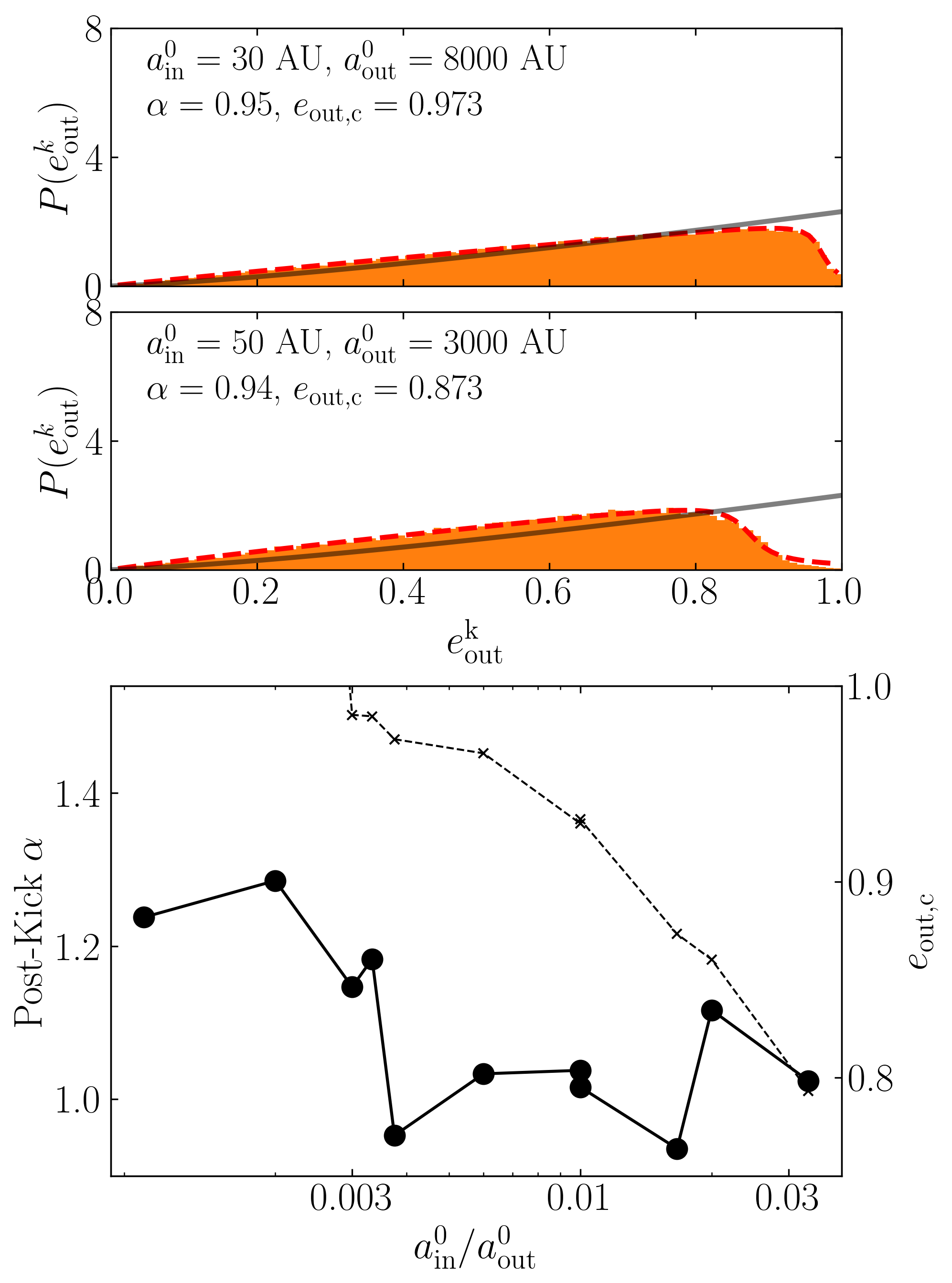}
    \caption{Similar to Fig.~\ref{fig:indicies} but for an initially mildly superthermal ($\alpha = 1.3$) distribution.}\label{fig:indicies_old}
\end{figure}

In Section~\ref{sec 2 3}, we considered the effect of mass loss on the distribution of the outer eccentricity for a highly superthermal distribution given by Eq.~\eqref{eq:pe}. In Figure~\ref{fig:indicies_old}, we also display the results for a weakly superthermal, $P(e_\OUT^0) \propto (e_\OUT^0)^{1.3}$ power-law distribution as found by \citet{hwang2022}.
It can be seen that the post-kick eccentricities of more hierarchical systems (smaller $a_{\rm in}^0 / a_{\rm out}^0$) are well-described by the same $\alpha \approx 1.3$ power law all the way out to very large $e_{\rm out}^k$.
On the other hand, the $e_{\rm out}^k$ distributions of less hierarchical systems have power-law indicies closer to $\alpha = 1$ and are truncated at modest values of $e_{\rm out, c} < 1$.
This result, in conjunction with the results shown in Figs.~\ref{fig:Survival fraction} and~\ref{fig:indicies}, further reinforce the conclusion of Section~\ref{sec 2 4} that the post-kick eccentricity distributions in very hierarchical systems are quite similar independent of the pre-kick eccentricity distribution, while the post-kick eccentricity distribution in mildly hierarchical systems better preserve the primordial $e_{\rm out}^0$ distribution.

\section{Black Hole Triple Inclination Distributions}\label{app:Idists}

\begin{figure*}
    \centering
    \includegraphics[width=0.7\textwidth]{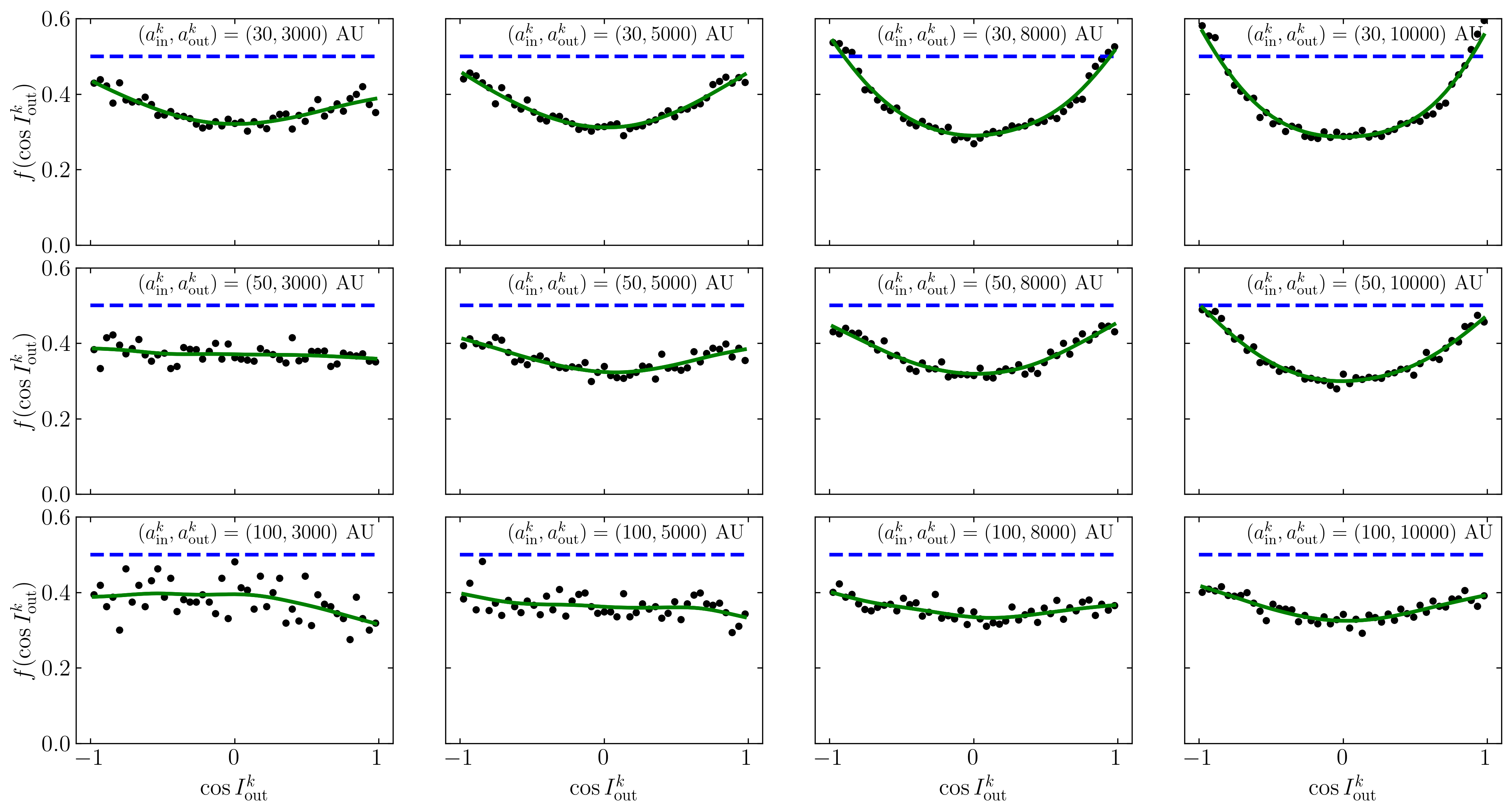}
    \caption{Black hole triple inclination distribution starting with a
    dynamically stable, isotropically oriented stellar triple for various
    orbital hierarchies labelled in the titles of subplots. Note that $f(\cos
    I_{\rm out}^k)$ is not normalized, and accounts for the probability that a
    stellar triple dissociates before becoming a stable hierarchical BH triple.
    The black line is taken from numerical simulations, the green line a
    smoothed version used to compute Fig.~\ref{fig:etamerge_Iconv_0}, and the
    black dashed line the initial $\cos I_{\rm out}^0$
    distribution.}\label{fig:Ihist_unif}
\end{figure*}

\begin{figure*}
    \centering
    \includegraphics[width=0.7\textwidth]{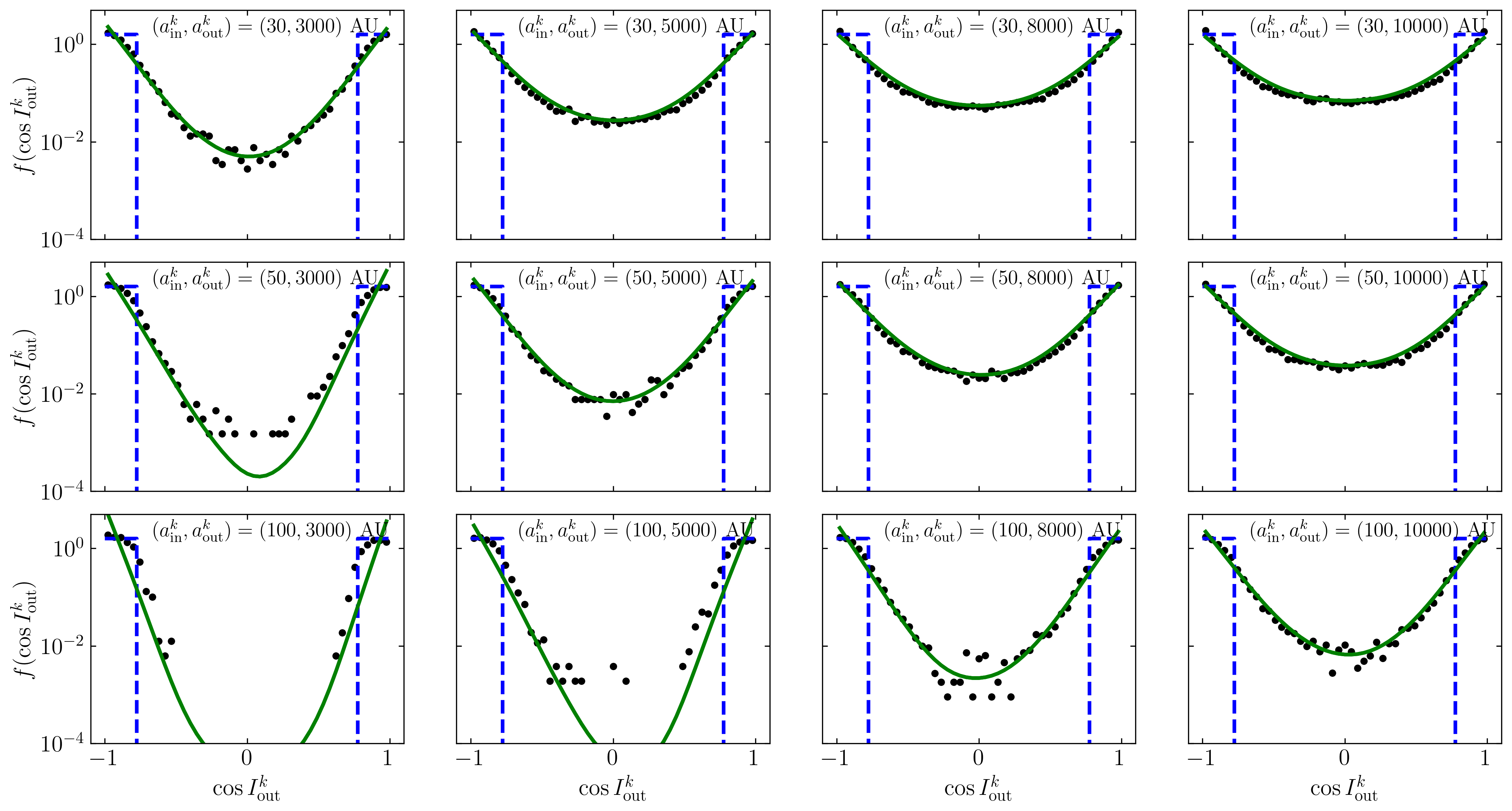}
    \caption{Same as Fig.~\ref{fig:Ihist_unif} but only for ZLK-stable stellar
    triples (note that the initial $I_{\rm out}^0$ distribution, the blue dashed
    line, is not uniform). This is used to compute
    Fig.~\ref{fig:etamerge_Iconv_lessinc}.}\label{fig:Ihist_lessinc}
\end{figure*}

In this section, we provide the set set of post-kick inclination distributions
(as seen in the bottom panels of Fig.~\ref{fig:Ihist2-unif}) for all pairs of
inner and outer semi-major axes: $a_{\rm in} \in [30, 50, 100]\;\mathrm{AU}$ and $a_{\rm
out} \in [3, 5, 8, 10] \times 10^3\;\mathrm{AU}$. We show these distributions
both for an initially randomly oriented stellar triple,
Fig.~\ref{fig:Ihist_unif} and for initally ZLK-stable stellar triples ($\cos
I_{\rm out}^0 \in [-1, -\sqrt{3/5}]\cup[\sqrt{3/5}, 1]$),
Fig.~\ref{fig:Ihist_lessinc}. Note that four of these panels are the same as the
lower two rows of Fig.~\ref{fig:Ihist2-unif}.

\end{document}